\documentclass[journal,10pt,twocolumn]{IEEEtran}\IEEEoverridecommandlockouts
\ifCLASSINFOpdf
\usepackage[pdftex]{graphicx}
\DeclareGraphicsExtensions{.pdf,.jpeg,.png} \else
\usepackage[dvips]{graphicx}
\DeclareGraphicsExtensions{.eps} \fi

\usepackage{cite}
\usepackage[normal]{caption2}\renewcommand{\figurename}{Fig.}
\usepackage[cmex10]{amsmath}
\usepackage{subfigure}

\begin{document}
\title{Multiple Watermarking Algorithm Based on Spread Transform Dither Modulation}

%\author{
%\IEEEauthorblockN{Xinchao~Li, Ju~Liu,~\IEEEmembership{Seniro Member,~IEEE,}
%        Jiande~Sun,~\IEEEmembership{Member,~IEEE,}, and Xiaohui~Yang}\\
%\IEEEauthorblockA{School of Information Science and Engineering,\\
%Shandong University\\
%Jinan 250100, P.R.China\\
%Email: juliu@sdu.edu.cn\\
%Tel: 86-531-88362323}
%}

\author{Xinchao~Li, Ju~Liu\IEEEauthorrefmark{1},~\IEEEmembership{Senior Member,~IEEE,}
        Jiande~Sun,~\IEEEmembership{Member,~IEEE,} Xiaohui~Yang, and Wei~Liu% <-this % stops a space

\thanks{Xinchao~Li, Ju~Liu, Jiande~Sun, and Xiaohui~Yang are with the School of Information Science and Engineering, Shandong University, Jinan,
250100, China (e-mail: juliu@sdu.edu.cn).}

\thanks{Ju~Liu and Wei Liu are with the Hisense State Key Laboratory of Digital Multi-Media Technology Co., Ltd, Qingdao, China.
}

\thanks{This work was supported partially by the National Basic Research Program of China (973 Program, No.2009CB320905), the National Natural Science Foundation of China (60872024), the Cultivation Fund of the Key Scientific and Technical Innovation Project (708059), Education Ministry of China for funding, Nature Science Foundation of Shandong Province (Q2008G03), Doctoral Program Foundation of Institutions of Higher Education of China (200804221023).
}

% <-this % stops a space
%\thanks{Manuscript received April 19, 2005; revised January 11, 2007.}}

}



%Telephone: (800) 555--1212, Fax: (888) 555--1212}


\maketitle

\begin{abstract}
%\boldmath
Multiple watermarking technique, embedding several watermarks in one carrier, has enabled many interesting applications. In this study, a novel multiple watermarking algorithm is proposed based on the spirit of spread transform dither modulation (STDM). It can embed multiple watermarks into the same region and the same transform domain of one image; meanwhile, the embedded watermarks can be extracted independently and blindly in the detector without any interference. Furthermore, to improve the fidelity of the watermarked image, the properties of the dither modulation quantizer and the proposed multiple watermarks embedding strategy are investigated, and two practical optimization methods are proposed. Finally, to enhance the application flexibility, an extension of the proposed algorithm is proposed which can sequentially embeds different watermarks into one image during each stage of its circulation. Compared with the pioneering multiple watermarking algorithms, the proposed one owns more flexibility in practical application and is more robust against distortion due to basic operations such as random noise, JPEG compression and valumetric scaling.

\end{abstract}



% \fi
\IEEEpeerreviewmaketitle

\section{Introduction}
\IEEEPARstart
In recent years, as the rapid development in the field of digital watermarking, multiple watermarking algorithms which give the possibility of embedding different watermarks in the same image, have received widespread attention since the pioneering contribution \cite{Cox97}, where the idea of embedding multiple watermarks in the same image is initially presented.
\par
Since then, multiple watermarking has enabled many interesting applications. In \cite{MW_ICASSP1999}, Mintzer and Braudaway suggest that the insertion of multiple watermarks can be exploited to convey multiple sets of information. Sencar and Memon \cite{MW_ambiguity} apply the selective detection of multiple embedded watermarks, which can yield lower false-positive rates compared with embedding a single watermark, to resist ambiguity attacks. Boato et al. \cite{MW_Tracing} introduce a new approach that allows the tracing and property sharing of image documents by sequentially embedding multiple watermarks into the data. Giakoumaki et al. \cite{MW_health} apply multiple watermarking algorithm to simultaneously addresses medical data protection, archiving, and retrieval, as well as source and data authentication.
\par
Meanwhile, different watermarking techniques and strategies have been proposed to achieve multiple watermarking. In \cite{On_multiple_watermarking}, Sheppard et al. discuss three methods to achieve multiple watermarking: rewatermarking, composite watermarking and segmented watermarking. Rewatermarking embeds watermarks one after another and the watermark signal could only be detected in the corresponding watermarked image using the former watermarked signal as the original image. The watermark embedded previously may be destroyed by the one embedded later. Composite watermarking discusses the extension of single watermarking algorithms to the case of multiple watermarking by introducing orthogonal watermarks \cite{MW_extention1,MW_extention2}. Being similar to these, CDMA based schemes \cite{MW_CDMA1,MW_CDMA2} use the orthogonal codes to modulate the watermarks from different users to derive the orthogonal watermarks. Unfortunately, they cannot guarantee the robustness in the case of blind extraction. Segmented watermarking embeds multiple watermarks into different segments of one image. Clearly, the number of segments limits the number and size of watermarks to be embedded \cite{Segment_WM}. The embedding pattern chosen for mapping watermarks to segments can greatly affect the robustness of each watermark against cropping attack \cite{Embedding_Patterns}.
\par
Other schemes embed different watermarks into different channels of the host data, e.g., different levels of wavelet transform coefficients\cite{MW_health}, or RGB of the color image \cite{MW_Phase-Modulation,MW_RGB}. In fact, the limited quantity of watermarks embedded would somehow constrain their application area.
\par
In this study, we focus on the techniques that can embed multiple watermarks into the same area and the same transform domain of one image, meanwhile, the embedded watermarks can be extracted independently and blindly in the detector without any interference.
\par
To this end, a novel multiple watermarking algorithm is proposed. It initially extends the spread transform dither modulation (STDM), a single watermarking algorithm, to the field of multiple watermarking. Moreover, through investigating the properties of the dither modulation (DM) quantizer and the proposed multiple watermarks embedding strategy, two optimization methods are presented which can improve the fidelity of the watermarked image significantly. Compared with the pioneering multiple watermarking algorithm \cite{MW_wang2003}, it has considerable advantages, especially in robustness against Gauss Noise, Salt\&Pepper Noise, JPEG Compression and Valumetric Scaling. Finally, some potential interesting applications are discussed and an application extension of our algorithm is proposed to realize image history management by sequentially embedded watermarks.
\par
The reminder of this paper is organized as follows. In section II, we briefly describe the main algorithm of spread transform dither modulation. In section III, the proposed multiple watermarking algorithm is introduced. In section IV, to improve the fidelity of the watermarked image, the properties of the dither modulation quantizer and the embedding strategy of the proposed algorithm are analyzed. In section V, two practical optimization methods are presented. In section VI, the efficiency of the two optimization methods is tested, meanwhile, the robustness of the proposed methods is assessed. Finally, some potential interesting applications of the proposed algorithm and the concluding remarks are summarized in section VII and VIII, respectively.

\section{Spread Transform Dither Modulation}
As the proposed multiple watermarking algorithm is based on Spread Transform Dither Modulation, a blind single watermarking algorithm belonging to the QIM family, introduction beginning with the basic QIM is appropriate.

\subsection{Quantization Index Modulation}
\begin{figure}[htb!]\label{fg1}
  \centering
  \scalebox{0.45}{\includegraphics[width=\textwidth]{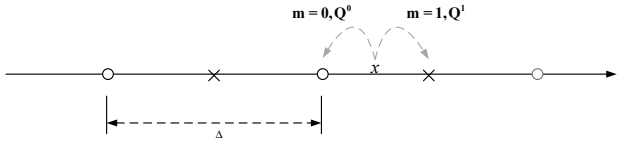}}
  \caption{Embedding one message bit, $m$, into one sample $x$ using original QIM, where sets of circles and crosses represent $\Omega^0$ and $\Omega^1$, respectively.}\label{QIM_traditional}
\end{figure}
\par
In the original QIM watermarking, a set of features extracted from the host signal are quantized by means of a quantizer chosen from a pool of predefined quantizers on the basis of the to-be-hidden message \cite{SW_QIM}. In the simplest case, a set of uniform quantizers are used leading to lattice-based QIM watermarking. As illustrated in Fig.\ref{QIM_traditional}, the basic QIM uses two quantizers $Q^0$ and $Q^1$ to implement the function, and each of them maps a value to the nearest point belonging to a class of predefined discontinuous points, one class ($\Omega^0$) represents bit ¡°0¡± while the other ($\Omega^1$) represents bit ¡°1¡± \cite{SW_QIM_Q.li_07Trans}. The standard quantization operation with step-size $\Delta$ is defined as
\begin{equation}\label{eq1}
%\small
\operatorname{Q}(x,\Delta)=\Delta \cdot \operatorname{round}(\frac{x}{\Delta})
\end{equation}
where the function round(.) denotes rounding a value to the nearest integer.
\par
In the embedding procedure, according to the message bit $m$, $Q^0$ or $Q^1$ is chosen to quantize the sample $x$ to the nearest quantization point $y$. For example, $Q^0$ and $Q^1$ may be chosen in such way that $Q^0$ quantizes x to even integers and $Q^1$ quantizes x to odd integers. If we wish to embed a ¡°0¡± bit, then $Q^0$ is chosen, else $Q^1$.
\par
In the detecting procedure, it is reasonable to assume the marked signal $y$ is corrupted by the attacker, resulting in a noisy signal $\tilde{y}$. The QIM detector is a minimum-distance decoder, which finds the quantization point closest to $\tilde{y}$ and outputs the estimated message bit $\tilde{m}$ \cite{Data-HidingCodes}.
\begin{equation}\label{eq2}
%\small
\tilde{m}=\operatorname*{argmin}\limits_{m \in \mathbf{0,1}}\operatorname{dist}(\tilde{y},\Omega^m)
\end{equation}
where ${\mathop{\rm dist}\nolimits} (\tilde{y},\Omega^m ) \buildrel \Delta \over = \mathop {\min }\limits_{s \in \Omega^m } \left| {\tilde{y} - s} \right|$.

\subsection{QIM-Dither Modulation}
Dither modulation, proposed by Chen and Wornell \cite{SW_QIM}, is an extension of the original QIM. Compared with the original QIM, it uses the pseudo-random dither signal, which can reduce quantization artifacts, to produce a perceptually superior quantized signal. Meanwhile, through the dither procedure, the quantization noise is independent from the host signal. The DM quantizer QDM is as following
\begin{equation}\label{eq3}
%\small
y=\operatorname{QDM(x,\Delta,d^m)}=\operatorname{Q}(x+d^m,\Delta)-d^m , m=0,1
\end{equation}
where $y$ is the marked signal of $x$ by DM quantizer, $d^m$ is the dither signal corresponding to the message bit $m$.
\begin{equation}\label{eq4}
%\small
d^1=d^0-\operatorname{sign}(d^0)\frac{\Delta}{2}
\end{equation}
where $d^0$ is a pseudo-random signal and is usually chosen with a uniform distribution over $[-\Delta/2,\Delta/2]$.
\par
In the detecting procedure, the detector firstly applies the QDM quantizer \eqref{eq3} to produce two signals $S^0$ and $S^1$, by embedding ``0" and ``1" into the received signal $\tilde{y}$ respectively.
\begin{equation}\label{eq5}
%\small
S^m=\operatorname{QDM}(\tilde{y},\Delta,d^m)=\operatorname{Q}(\tilde{y}+d^m,\Delta)-d^m , m=0,1
\end{equation}
where $d^m$ must be exactly the same as which in the embedding procedure. Note that the pseudo-random signal $d^0$ can be considered as a key to improve the security of the system, and in what follows, this secret signal is referenced as the dither factor, $df$.
\par
The detected message bit is then estimated by judging which of these two signals has the minimum Euclidean distance to the received signal $\tilde{y}$, in the same manner as \eqref{eq2}.
\begin{equation}\label{eq6}
%\small
\tilde{m}=\operatorname*{argmin}\limits_{m \in \mathbf{0,1}}\operatorname{dist}(\tilde{y},S^m)
\end{equation}

\subsection{QIM-Spread Transform Dither Modulation}

As an important extension of the original QIM, STDM applies the idea of projection modulation. It utilizes the DM quantizer to modulate the projection of the host vector along a given direction. This scheme combines the effectiveness of QIM and robustness of spread-spectrum system, and provides significant improvements compared with DM.

\begin{figure}[htb!]
  \centering
  \scalebox{0.45}{\includegraphics[width=\textwidth]{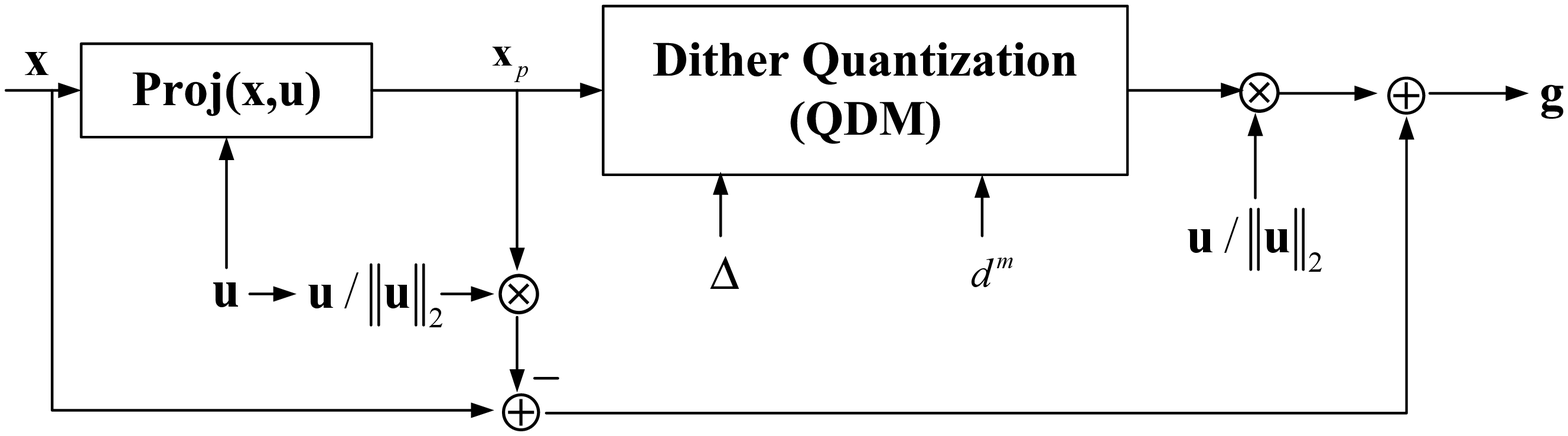}}
  \caption{Block diagram of spread transform dither modulation}\label{QIM_STDM}
\end{figure}
To embed one message bit $m$, a host vector \textbf{x}, consisting of samples to be embedded, is projected onto a random vector \textbf{u} to get the projection $x_p$. Then, the projection $x_p$ is modulated according to the message bit $m$ using the DM quantizer \eqref{eq3}. This procedure can be illustrated in Fig.\ref{QIM_STDM}, and the watermarked vector \textbf{g} is as follows,
\begin{equation}\label{eq7}
%\small
{\bf{g}} = {\bf{x}} + (\frac{{{\mathop{\rm QDM}\nolimits} ({\mathop{\rm proj}\nolimits} ({\bf{x}},{\bf{u}}),\Delta ,d^m ) - {\mathop{\rm proj}\nolimits} ({\bf{x}},{\bf{u}})}}{{\left\| {\bf{u}} \right\|_2 }}){\bf{u}}
\end{equation}
where ${\mathop{\rm proj}\nolimits} ({\bf{x}},{\bf{u}}) \buildrel \Delta \over = \frac{{\left\langle {{\bf{x}},{\bf{u}}} \right\rangle }}{{\left\| {\bf{u}} \right\|_2 }}$, $\left\langle {{\bf{x}},{\bf{u}}} \right\rangle$ is the inner product of \textbf{x} and \textbf{u},
$\left\| {\; \cdot \;} \right\|_2$ denotes the $L^2$-norm operation. $\Delta$ is the quantization step generated from a pseudo-random generator.
\par
In the detecting procedure, the detector projects the received vector ${\bf{\tilde g}}$ onto the random vector \textbf{u}. And then, it utilizes the DM detector to estimate the message bit $\tilde{m}$ from the projection, in the same manner as \eqref{eq5} and \eqref{eq6}. This can be expressed as follows,
\begin{equation}\label{eq8}
%\small
\tilde m = \mathop {\arg \min }\limits_{m \in \{ 0,1\} } {\mathop{\rm dist}\nolimits} ({\mathop{\rm proj}\nolimits} ({\bf{\tilde g}},{\bf{u}}),{\mathop{\rm QDM}\nolimits} ({\mathop{\rm proj}\nolimits} ({\bf{\tilde g}},{\bf{u}}),\Delta ,d^m )\;)
\end{equation}
Note that, the random vector \textbf{u} and the random positive real number $\Delta$ used in the STDM detector must be exactly the same as they are in the embedder, and can be considered as two keys which are only known to the embedder and detector, thereby improving the security of the system.

\section{Multiple Watermarking Algorithm}
Based on the algorithms mentioned above, we extend the spread transform dither modulation (STDM), a single watermarking algorithm, to the field of multiple watermarking application. The proposed multiple watermarking algorithm, namely STDM-Multiple Watermarking (STDM-MW), can embed multiple watermarks into the same area and the same transform domain of one image, meanwhile, the embedded watermarks can be extracted independently and blindly in the detector without any interference.

\subsection{Fundamental Idea }
As mentioned in section II, to embed a single message bit, $m$, STDM modulates the projection of the host vector $\bf{x}$ along a given direction $\bf{u}$. The modulated host vector $\bf{g}$ can be expressed as follows,
\begin{equation}\label{eq11}
%\small
{\bf{g}} = {\bf{x}} + k{\bf{u}}
\end{equation}
\par
To detect the message bit, the detector projects the modulated vector $\bf{g}$ onto the given direction $\bf{u}$. And then, it utilizes the DM detector to estimate the message bit from the projection. This detection mechanism induces the vector $\bf{g}$ must be subject to
\begin{equation}\label{eq12}
%\small
{\mathop{\rm proj}\nolimits} ({\bf{g}},{\bf{u}}) = {\mathop{\rm QDM}\nolimits} ({\mathop{\rm proj}\nolimits} ({\bf{x}},{\bf{u}}),\Delta ,d^m )
\end{equation}
\par
Thus, the embedding procedure is actually to derive the scaling factor $k$ used in \eqref{eq11} to make the modulated vector $\bf{g}$ in the form of \eqref{eq12}. Substituting \eqref{eq11} into \eqref{eq12}, the scaling factor $k$ can be given by
\begin{equation}\label{eq13}
%\small
k  = \frac{{{\mathop{\rm QDM}\nolimits} ({\mathop{\rm proj}\nolimits} ({\bf{x}},{\bf{u}}),\Delta ,d^m ) - {\mathop{\rm proj}\nolimits} ({\bf{x}},{\bf{u}})}}{{\left\| {\bf{u}} \right\|_2 }}
\end{equation}
\par
Inspired by this, to embed multiple message bits, $m_1$, $m_2$,..., $m_n$, into the same host vector $\bf{x}$, we can modulate the projection of the host vector $\bf{x}$ along different given directions, $\bf{u}_1$, $\bf{u}_2$,..., $\bf{u}_n$. The modulated host vector $\bf{g}$ can be expressed as follows
\begin{equation}\label{eq14}
%%{\bf{g}} = {\bf{x}} + k_1{\bf{u}_1}+ k_2{\bf{u}_2}+...+ k_n{\bf{u}_n}
%\small
\;{\bf{g}} = {\bf{x}} + {\bf{U}}{\bf{K}}
\end{equation}
where ${\bf{U}} = [{\bf{u}}_1 ,{\bf{u}}_2 ,...,{\bf{u}}_n ]$, ${\bf{K}} = [k_1 ,k_2 ,...,k_n ]^T$.

\par
To detect the message bits, the modulated vector $\bf{g}$ is projected onto the given directions, $\bf{u}_1$, $\bf{u}_2$,..., $\bf{u}_n$, respectively. And then, the DM detector is used to estimate each message bit from the corresponding projection. Thus, in the same manner as \eqref{eq12}, the modulated vector $\bf{g}$ must be subject to the following equation,

\begin{equation}\label{eq15}
%\small
\begin{array}{l}
 \left\{ \begin{array}{l}
 {\mathop{\rm proj}\nolimits} ({\bf{g}},{\bf{u}}_1 ) = {\mathop{\rm QDM}\nolimits} ({\mathop{\rm proj}\nolimits} ({\bf{x}},{\bf{u}}_1 ),\Delta_1 ,d_{1}^{m_1} ) \\
 {\mathop{\rm proj}\nolimits} ({\bf{g}},{\bf{u}}_2 ) = {\mathop{\rm QDM}\nolimits} ({\mathop{\rm proj}\nolimits} ({\bf{x}},{\bf{u}}_2 ),\Delta_2 ,d_{2}^{m_2} ) \\
 .........\;......... \\
 {\mathop{\rm proj}\nolimits} ({\bf{g}},{\bf{u}}_n ) = {\mathop{\rm QDM}\nolimits} ({\mathop{\rm proj}\nolimits} ({\bf{x}},{\bf{u}}_n ),\Delta_n ,d_{n}^{m_n} ) \\
 \end{array} \right. \\
  \\
 \end{array}
\end{equation}
where $d_{j}^{m_j}$ is the dither signal in the direction ${\bf{u}_j}$ corresponding to the message bit $m_j$.
\par
By substituting \eqref{eq14} into \eqref{eq15}, n equations can be obtained. These are expressed as follows in the matrix form,
\begin{equation}\label{eq16}
%\small
{\bf{U}}_{\bf{I}} {\bf{K}} = {\bf{QDMV}}-{\bf{P}}
\end{equation}
where
\par
$\begin{array}{l}
 \;\;\;\;\;\;\;\;\;\;\;{\bf{U}}_{\bf{I}}  = \Lambda _U {\bf{U}^T}{\bf{U}} ,\;\Lambda _U  = [\frac{1}{{\left\| {{\bf{u}}_1 } \right\|}},\frac{1}{{\left\| {{\bf{u}}_2 } \right\|}},...,\frac{1}{{\left\| {{\bf{u}}_n } \right\|}}] \\
 \;\;\;\;\;\;\;\;\;\;\;{\bf{P}} = [{\mathop{\rm proj}\nolimits} ({\bf{x}},{\bf{u}}_1 ),{\mathop{\rm proj}\nolimits} ({\bf{x}},{\bf{u}}_2 ),...,{\mathop{\rm proj}\nolimits} ({\bf{x}},{\bf{u}}_n )]^T  \\
 \;\;\;\;\;\;\;\;\;\;\;{\bf{QDMV}} = [QDMV_1 ,QDMV_2 ,...,QDMV_n ]^T  \\
 \;\;\;\;\;\;\;\;\;\;\;QDMV_j  = {\mathop{\rm QDM}\nolimits} ({\mathop{\rm proj}\nolimits} ({\bf{x}},{\bf{u}}_j ),\Delta_j ,d_{j}^{m_j} )\\
 \end{array}$
 \par
 $\;$
 \par
 From \eqref{eq16}, the scaling factor sequence $\bf{K}$ can be calculated by
\begin{equation}\label{eq17}
%\small
{\bf{K}} = {\bf{U}}_{_{\bf{I}} }^{^{ - 1} } ({\bf{QDMV}} - {\bf{P}})
\end{equation}
\par
Finally, according to \eqref{eq14}, the watermarked host vector $\bf{g}$ which carries $n$ message bits can be generated. Note that, to make \eqref{eq17} tenable, the length of the host vector $\bf{x}$, namely $L$, must be no less than the number of embedded message bits, $n$, i.e., $L \ge n$, (see Appendix A).
\par
In the detecting procedure, we can apply the STDM detector \eqref{eq8} to estimate every single bit $\widetilde m_j$ from the projection of the received vector ${\bf{\tilde g}}$ along the corresponding direction ${\bf{u}}_j$, independently. This can be expressed as follows,

\begin{equation}\label{eq18}
%\footnotesize
\begin{array}{l}
\widetilde m_j  = \mathop {\arg \min }\limits_{m_j \in \{ 0,1\} } {\mathop{\rm dist}\nolimits} ({\mathop{\rm proj}\nolimits} ({\bf{\tilde g}},{\bf{u}}_j ),{\mathop{\rm QDM}\nolimits} ({\mathop{\rm proj}\nolimits} ({\bf{\tilde g}},{\bf{u}}_j ),\Delta _j ,d_{j}^{m_j} ),\\
\;\;\;\;\;\;\;\;\;\;\;\;\;\;\;\;\;\;\;\;\;\;\;\;\;\;\;\;\;\;\;\;\;\;\;\;\;\;\;\;\;\;\;\;\;\;\;\;\;\;\;\;\;\;\;\;\;\;\;\;\;\;j=1,2,...,n
\end{array}
\end{equation}

\subsection{Detailed Implementation }
\begin{figure*}[htb!]
\begin{center}
  \scalebox{0.9}{\includegraphics[width=\textwidth]{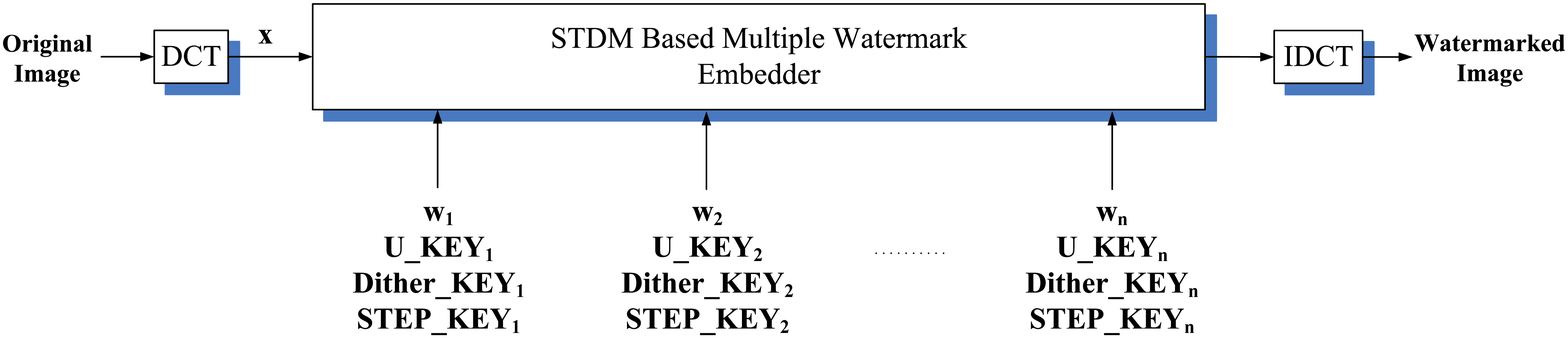}}
  \caption{Block diagram of STDM-Multiple Watermarking embedder}\label{STDM_mutiple}
\end{center}
\end{figure*}

\begin{figure*}[htb!]
\begin{center}
  \scalebox{0.7}{\includegraphics[width=\textwidth]{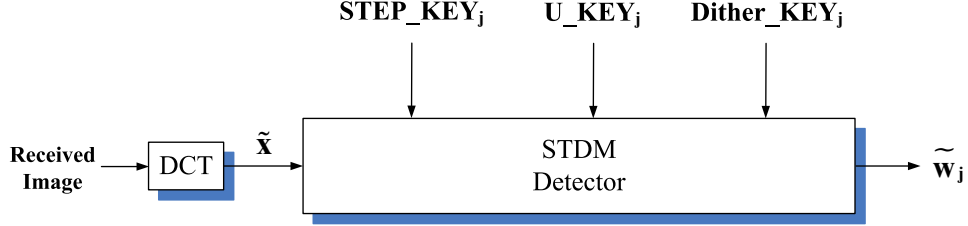}}
  \caption{Block diagram of STDM-Multiple Watermarking detector for the $j^{th}$ user}\label{STDM_mutiple_detector}
\end{center}
\end{figure*}

As illustrated in Fig.\ref{STDM_mutiple} and Fig.\ref{STDM_mutiple_detector}, the proposed scheme, STDM-MW, consists of two parts, the embedder (Fig.\ref{STDM_mutiple}) and the detector (Fig.\ref{STDM_mutiple_detector}). In this scheme, each user is given three secret keys, $STEP\_KEY$, $U\_KEY$ and $Dither\_KEY$, to implement watermark embedding and detecting. It is assumed that there are $n$ users and the watermark sequence of the $j^{th}$ user is $\bf{w}_j$, ${\bf{w}_j} = [w_{j1} ,w_{j2} ,...,w_{jN} ]$, with length $N$.

\par
The embedding procedure is as follows,
\par
\begin{figure}[htb!]
\begin{center}
  \scalebox{0.4}{\includegraphics[width=\textwidth]{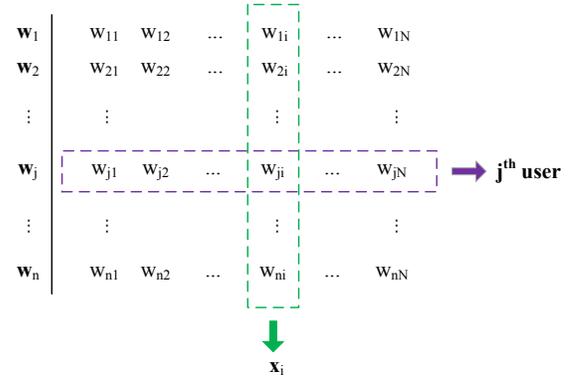}}
  \caption{Parameters arrangement, the arrangement for projective vector ${\bf{u}}$, dither factor $df$ and step size $\Delta$ is the same as it is for watermark $w$.}\label{parameter_arrangement}
\end{center}
\end{figure}

\hangafter=1\hangindent=1.5em\noindent
(a) Divide the image into disjoint $8 \times 8$ blocks of pixels, and perform DCT transform to each block to gain its DCT coefficients. A part of these coefficients will be selected to form a single vector, denoted as the host vector ${\bf{x}}_i(i=1,2,...,N)$, ${\bf{x}}_i = [x_1 ,x_2 ,...,x_L ]$, with length L. As illustrated in Fig.\ref{parameter_arrangement}, each host vector ${\bf{x}}_i$ is used to embed one bit sequence $[w_{1i},w_{2i},...,w_{ni}]$, the $j^{th}$ element of which is corresponding to the $j^{th}$ user's $i^{th}$ bit.
\par
\hangafter=1\hangindent=1.5em\noindent
(b) Use the secret keys, $STEP\_KEY$, $U\_KEY$ and $Dither\_KEY$, of each user to generate the step sizes $\Delta_{ji}$, the random projective vectors ${\bf{u}}_{ji}$ and the dither factors $df_{ji}$ for each host vector ${\bf{x}}_i$, respectively. According to the message bit $w_{ji}$, the final dither signal $d_{ji}^{w_{ji}}$ can be generated using $df_{ji}$.
\par
\hangafter=1\hangindent=1.5em\noindent
(c) Embed each bit sequence $[w_{1i},w_{2i},...,w_{ni}]$ by modulating each host vector ${\bf{x}}_i$ into ${\bf{g}}_i$ using the method mentioned in III-A , based on the parameters, $[{\bf{u}}_{1i},{\bf{u}}_{2i},...,{\bf{u}}_{ni}]$, $[d_{1i}^{w_{1i}},d_{2i}^{w_{2i}},...,d_{ni}^{w_{ni}}]$, $[\Delta _{1i},\Delta _{2i},...,\Delta _{ni}]$, calculated in step (b). Finally, transform the modified coefficients back to form the watermarked image.

\par
During the transmission, the watermarked image may sustain certain attacks, intentional or unintentional, and become a distorted image at the receiver. Each user can use his own secret keys to detect his own watermark independently.
\par
The detecting procedure of the $j^{th}$ user is as follows
\par
\hangafter=1\hangindent=1.5em\noindent
(a) Form each host vector ${\bf{\tilde g}}_i$ of the received image in the same manner as step (a) in the embedding procedure.
\par
\hangafter=1\hangindent=1.5em\noindent
(b) use the secret keys, $STEP\_KEY_j$, $U\_KEY_j$ and $Dither\_KEY_j$, of the $j^{th}$ user to generate the step sizes $[\Delta_{j1},\Delta_{j2},...,\Delta_{jN}]$, the random projective vectors $[{\bf{u}}_{j1},{\bf{u}}_{j2},...,{\bf{u}}_{jN}]$ and the dither factors $[df_{j1},df_{j2},...,df_{jN}]$, respectively.
\par
\hangafter=1\hangindent=1.5em\noindent
(c) Use the STDM detector to detect every bit $\tilde{w}_{ji}$ from each host vector ${\bf{\tilde x}}_i$, based on the parameters, ${\bf{u}}_{ji}$, $df_{ji}$ and $\Delta _{ji}$.

Note that, with an eye to the robustness of STDM-MW against valumetric scaling, the step-size $\Delta$ should be multiplied by the mean intensity of the whole image.

\section{Analysis of STDM-Multiple Watermarking}
Through experiment, it is found that along with the increase of the number of watermarks embedded, the quality of the images declines in vary degrees. To address this issue, further analysis of the embedding strategy of STDM-Multiple Watermarking is demanded.
\par
As is widely known, in the case of Imperceptible $\&$ Robust watermarking, owning the same robustness, the more imperceptible, the more effective the algorithm is. In most cases, the imperceptibility of the watermark, in other words the fidelity of the watermarked image, is measured in PSNR, which varies inversely with the mean squared error, MSE. Referencing Appendix B, we have
\begin{equation}\label{eq19}
%\small
MSE \propto \left\| {\bf{C'} - \bf{C}} \right\|_2
\end{equation}
where $\bf{C}$ and $\bf{C}'$ are the DCT coefficient vectors of the original image and the watermarked one.
\par
Thus, under the PSNR measurement, the smaller the Euclidian distance between the watermarked coefficient and the original one is, the higher the fidelity of the watermarked image will be. According to this idea, to improve the fidelity of the watermarked image, we need to produce the watermarked vector that is closest to the host vector.
\par
At the very beginning, as the embedding procedure of STDM-Multiple Watermarking is based on Dither Modulation, it is appropriate to investigate the DM quantizer in a deeper way.

\par

\subsection{Dither Modulation Based Single Watermarking }
From section II-B, to embed one message bit $m$, the original DM quantizer, QDM, quantizes the point $x$ to ($\Delta round(\frac{{x + d^m }}{\Delta }) - d^m$). However, ignoring the imperceptible constraint (minimum Euclidian distance), we can quantize the point $x$ to any point $b_i$, $b_i \in {\bf{B}}$.
\begin{equation}\label{eq20}
%\small
{\bf{B}} = \{ b|b = \beta \Delta  - d^m ,\;\beta  \in Z\}
\end{equation}
\par
Definitely, any points in ${\bf{B}}$ have the same detection robustness according to the DM detection mechanism, \eqref{eq5} and \eqref{eq6}. In what follows, this kind of points are defined as the DM quantization points of point $x$.
\par

\begin{figure}[htb!]
\begin{center}
  \scalebox{0.45}{\includegraphics[width=\textwidth]{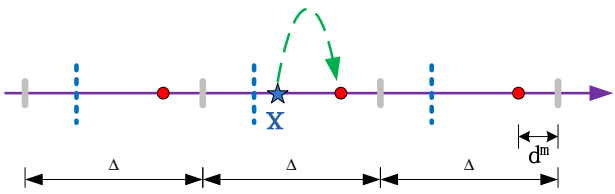}}
  \caption{Utilizing DM to embed one message bit $m$ into point $x$, where the set of circles represents quantization points in $B$, (assuming $d^m>0$). Dotted-lines, $L=\{l|l=((2\alpha  + 1)\frac{\Delta }{2} - d^m,\;\alpha \in Z\}$, denote the median point between two adjacent quantization points.}\label{DM_embedder}
\end{center}
\end{figure}

As illustrated in Fig.\ref{DM_embedder}, in the case of DM single watermarking, it is optimal to use \eqref{eq3}, which is equivalent to $\beta=round(\frac{{x + d^m }}{\Delta })$ in \eqref{eq20}, to choose the final quantization point, because the selected one is the closest point to $x$ among all the DM quantization points of $x$,(i.e., points in $B$).

\par
\begin{figure}[htb!]
\begin{center}
  \scalebox{0.45}{\includegraphics[width=\textwidth]{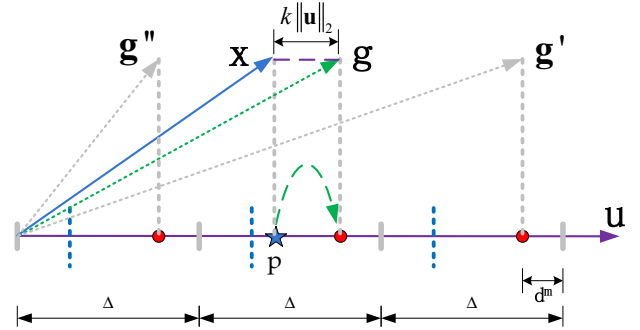}}
  \caption{Utilizing STDM to embed one message bit $m$ into one host vector $\bf{x}$, where $\bf{u}$ is the projective vector and $\bf{g}$ is the watermarked vector. The set of circles represents the DM quantization points of the projection point $p$ of the host vector $\bf{x}$ along the direction $\bf{u}$.}\label{STDM_embedder_explain}
\end{center}
\end{figure}

Inspired by this idea, in the original STDM, as illustrated in Fig.\ref{STDM_embedder_explain}, we can modulate the host vector $\bf{x}$ to any vector (${\bf{g}}''$,${\bf{g}}'$,${\bf{g}}$), whose projection point is the DM quantization point of the host vector's projection point $p$.
\par
However, the imperceptible constraint must be considered. Referencing \eqref{eq11}, the Euclidian distance $dis\_v$ between the watermarked vector $\bf{g}$ and the host vector $\bf{x}$, is proportional to $k$, which is actually the distance $dis\_p$ between the host vector's projection point $p$ and $p$'s DM quantization point. This can be formulated as follows,
\begin{equation}\label{eq21}
%\small
dis\_v=\left\| {{\bf{g}} - {\bf{x}}} \right\|_2  = \left\| {{\bf{x}} + k{\bf{u}} - {\bf{x}}} \right\|_2  = k\left\| {\bf{u}} \right\|_2=dis\_p
\end{equation}
%%where $dis\_p$ denotes the distance between the projection of the host vector and its DM quantization point.
\par
As DM quantizer \eqref{eq3} can generate the quantization point that is closest to the original point, it can find the closest DM quantization point to the host vector's projection point, i.e., the DM quantizer can make $dis\_p$ minimum. Thus, it is optimal to use DM quantizer to modulate the host vector $\bf{x}$ to vector $\bf{g}$ by \eqref{eq7}. In this way, the minimum $dis\_v$ can be guaranteed.

\subsection{Embedding Strategy of STDM-Multiple Watermarking}
As mentioned above, DM quantizer is optimal for STDM in the case of single watermarking. Unfortunately, it seems that this strategy is not optimal in the case of multiple watermarking.
\par
As mentioned in III-A, in the case of multiple watermarking, if $n$ message bits are embedded, the host vector $\bf{x}$ must be modulated along $n$ given directions to form the watermarked vector $\bf{g}$. For each direction, the projection of the watermarked vector $\bf{g}$ must be the closet DM quantization point to the host vector $\bf{x}$'s projection point.

\par
\begin{figure}[htb!]
\begin{center}
  \scalebox{0.45}{\includegraphics[width=\textwidth]{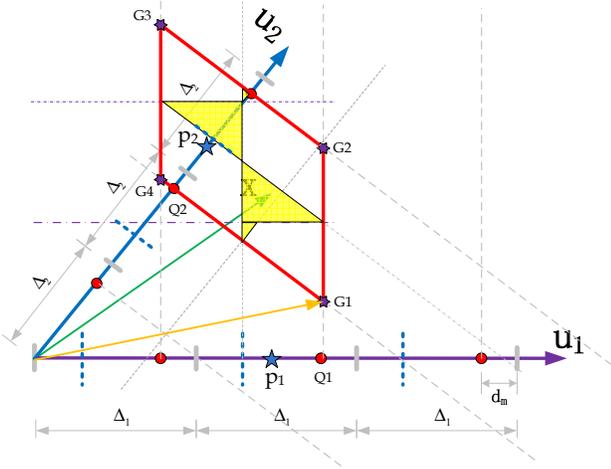}}
  \caption{Utilizing the original embedding strategy STDM-MW to embed two message bits into one host vector $\bf{x}$, where $\bf{u}_1$ and $\bf{u}_2$ are the two projective vectors denoting the quantization directions. $p_1$ and $p_2$ are the projection points of $\bf{x}$ along $\bf{u}_1$ and $\bf{u}_2$, respectively. The circles along $\bf{u}_1$ and $\bf{u}_2$ denote the DM quantization points, belonging to the point set ${\bf{B}}_1$ and ${\bf{B}}_2$,respectively. ${\bf{B}}_j=\{ b|b =\beta_j \Delta_j  - d_j^{m_j} ,\;\beta_j  \in Z \}$. }\label{un_optimal}
\end{center}
\end{figure}

As illustrated in Fig.\ref{un_optimal}, it is a simple example for two users, that is embedding two bits into the host vector $\bf{x}$. To do this, host vector $\bf{x}$ must be projected along the projective vectors $\bf{u}_1$ and $\bf{u}_2$ to gain the projection points $p_1$ and $p_2$, respectively. And then, points $p_1$ and $p_2$ are quantized into their closet DM quantization points, $Q_1$ and $Q_2$, respectively. Finally, host vector $\bf{x}$ is modulated into vector $\bf{G}_1$.

\par
However, this original embedding strategy, using the closest DM quantization point as the final quantization point of the projection point, can not product the closet watermarked vector to the host vector. Actually, vectors $\bf{G}_1$, $\bf{G}_2$, $\bf{G}_3$ and $\bf{G}_4$ can all be selected as the watermarked vector of the host vector $\bf{x}$ while owning the same detection robustness. And, as shown in Fig.\ref{un_optimal}, vector $\bf{G}_1$, the original selected one, dose not have the minimum Euclidian distance to the host vector $\bf{x}$ among the four alternative ones. In practice, vector $\bf{G}_2$ is the closest one.
\par
Thus, it is not optimal to use vector $\bf{G}_1$ to play as the watermarked vector. More specifically, once the host vector $\bf{x}$ belongs to the shadowed area in the parallelogram in Fig.\ref{un_optimal}, it is not optimal to use the original embedding strategy to select the quantization point along each direction and generate the watermarked vector.
\par
The original multiple watermarks embedding strategy \eqref{eq15} and \eqref{eq17} must be rewritten as
\begin{equation}\label{eq22}
%\small
\left\{ \begin{array}{l}
 {\mathop{\rm proj}\nolimits} ({\bf{g}},{\bf{u}}_1 ) = Qp_1  \\
 {\mathop{\rm proj}\nolimits} ({\bf{g}},{\bf{u}}_2 ) = Qp_2  \\
 ...........\;.......... \\
 {\mathop{\rm proj}\nolimits} ({\bf{g}},{\bf{u}}_n ) = Qp_n  \\
 \end{array} \right.
\end{equation}

 \begin{equation}\label{eq23}
 %\small
{\bf{K}} = {\bf{U}}_{_{\bf{I}} }^{^{ - 1} } ({\bf{Qp}} - {\bf{P}})
\end{equation}
where $Qp_j$ denotes one DM quantization point in the j-th direction,
%\par
%$\;$
%\par

$
\begin{array}{l}
 \;\;\;\;\;\;\;\;\;\;\;{\bf{Qp}} = [Qp_1 ,Qp_2 ,...,Qp_n ]^T ,\; \\
 \;\;\;\;\;\;\;\;\;\;\;Qp_j  \in {\bf{B}}_j, {\bf{B}}_j=\{ b|b = \beta _j \Delta _j  - d_j^{m_j} ,\beta _j  \in {\rm Z}\}  \\
 \end{array}
$

\par
$\;$
\par

Substituting \eqref{eq23} into \eqref{eq14}, the watermarked vector can be given by

\begin{equation}\label{eq105}
%%{\bf{g}} = {\bf{x}} + k_1{\bf{u}_1}+ k_2{\bf{u}_2}+...+ k_n{\bf{u}_n}
%\small
\;{\bf{g}} = {\bf{x}} + {\bf{U}}{\bf{U}}_{_{\bf{I}} }^{^{ - 1} } ({\bf{Qp}} - {\bf{P}})
\end{equation}

As there are many DM quantization points in each direction, there are several combinations to make $\bf{Qp}$. This will form a vector pool for $\bf{Qp}$, namely $\bf{Qp\_S}$. Vectors in $\bf{Qp\_S}$ can all be chosen as $\bf{Qp}$ in \eqref{eq105}, and correspondingly, a vector pool for the watermarked vector ${\bf{g}}$ is generated, namely {\bf{g\_S}}. The goal of our optimization procedure is to find the closest one to the host vector ${\bf{x}}$ from this vector pool $\bf{g\_S}$, and finally use this vector to play as the optimized watermarked vector.

\section{Optimization for STDM-Multiple Watermarking}

As mentioned above, obviously, if all the candidate vectors in the pool $\bf{g\_S}$ are traversed, the one which is closest to the host vector will be found ultimately. However, as the infinite size of $\bf{g\_S}$, this procedure is not practical. To address this issue, the optimization procedure is divided into two cases, the special case and the general case.

\subsection{Special Case: Multiple Watermarking using Orthogonal Projective Vectors}
It has been observed that the goal of our optimization procedure is to find the closet watermarked vector to the host vector, i.e., the Euclidian distance between them is minimum. According to \eqref{eq105}, the Euclidian distance, $dis\_v$, can be expressed as follows,
\begin{equation}\label{eq27}
%\small
\begin{array}{l}
 dis\_v = \left\| {{\bf{g}} - {\bf{x}}} \right\|_2  = \sqrt {({\bf{Qp}}- {\bf{P}})^T {\bf{U}}_e ({\bf{Qp}} - {\bf{P}})}
 \end{array}
\end{equation}
where ${\bf{U}}_e  = \Lambda _U ^{ - 1} ({\bf{U}}^T {\bf{U}})^{ - 1} \Lambda _U ^{ - 1}$.
\par
If the projective vectors $\bf{u_1}$,$\bf{u_2}$,...,$\bf{u_n}$ are preprocessed by Gram-Schmidt orthogonalization, the matrix ${\bf{U}}_e$ will be Identity matrix ${\bf{I}}_n$, and $dis\_v$ is actually the Euclidian distance between the vector of DM quantization points, $\bf{Qp}$, and the vector of projection points, $\bf{P}$.
\begin{equation}\label{eq36}
%\small
dis\_v=\left\| {{\bf{Qp}} - {\bf{P}}} \right\|_2=\sqrt {\sum\limits_j {({\bf{Qp}}(j) - {\bf{P}}(j))^2 } }
\end{equation}

As QDM quantizer \eqref{eq3} can minimize each item in \eqref{eq36}, the original embedding strategy, using the closest DM quantization point as the final quantization point of the projection point, is optimal in the case of multiple watermarking using orthogonal projective vectors. Note that, in the following description, this special case will be referred as STDM-MW-Uorth.
\par
\begin{figure}[htb!]
\begin{center}
  \scalebox{0.45}{\includegraphics[width=\textwidth]{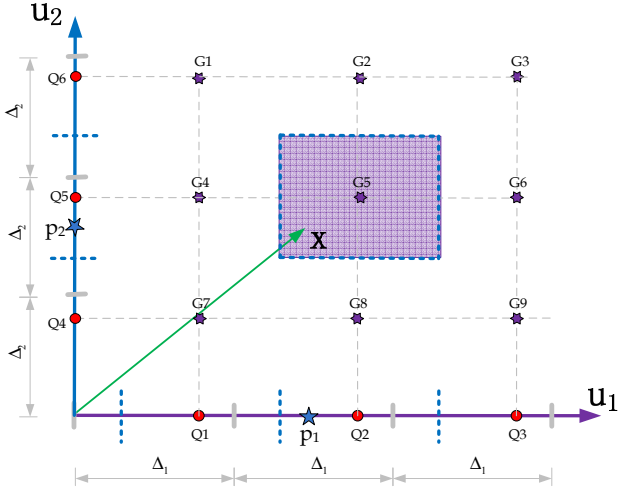}}
  \caption{Embedding two message bits into one host vector $\bf{x}$ in the case of $\bf{u_1}$ and $\bf{u_2}$ are orthogonal projective vectors, and ${\bf{G}}_5$ is the optimal watermarked vector for ${\bf{x}}$.}\label{STDM_MW_Uorth}
\end{center}
\end{figure}

The simple example for this case is illustrated in Fig.\ref{STDM_MW_Uorth}, if the host vector $\bf{x}$ belongs to the rectangle area centered by ${\bf{G}}_i$ with width $\Delta_1$ and height $\Delta_2$, it will be modulated to the vector ${\bf{G}}_i$. Obviously, ${\bf{G}}_i$ is the optimal watermarked vector for ${\bf{x}}$.

\subsection{General Case: Multiple Watermarking using Unorthogonal Projective Vectors}
In general, it is not realistic to expect the projective vectors $\bf{u_1}$,$\bf{u_2}$,...,$\bf{u_n}$ are orthogonal with each other. Thus, taking a tradeoff between PSNR and time efficiency, we propose two methods for the general case to find the optimized watermarked vector which is much closer to the host vector, namely STDM-MW-Poptim and STDM-MW-Qoptim.

\subsubsection{STDM-MW-Poptim}
In STDM-MW-Poptim, along each direction, $t$ quantization points, which are near the projection point of the host vector, are selected to form the point-set for this direction. This can be expressed as follows
\begin{equation}\label{eq24}
%\small
{\bf{H}}_j  = \{ h|h = \Delta _j (floor(\frac{x}{{\Delta _j }}) + k) - d_j^{m_j} ,\;\;k \in {\rm Z}\}
\end{equation}
where ${\bf{H}}_j$ denotes the point-set of the j-th direction.
\par
And then, one point of each point-set is selected to form a vector $\bf{PoQp}$. It can be used to substitute the vector $\bf{Qp}$ in \eqref{eq105}, and the watermarked vector can be calculated by
\begin{equation}\label{eq25}
%\small
{\bf{g}}_i  = {\bf{x}} + {\bf{UU}}_{\bf{I}} ^{ - 1} ({\bf{PoQp}}_i  - {\bf{P}}),\;\;i=1,2,...,F
\end{equation}
where, assuming there are $n$ bits to be embedded in one host vector, in other words $n$ quantization directions are given for one host vector, thus there are $F=t^n$ ways to choose one element from $n$ point-sets (of length $t$) to form the vector $\bf{PoQp}$. And correspondingly, $F$ watermarked vectors $\bf{g}$ are produced.
\par
The final optimized watermarked vector ${\bf{g}}_{optim}$ is then given by judging which of these watermarked vectors produced in \eqref{eq25} has the minimum Euclidean distance to the host vector $\bf{x}$.
\begin{equation}\label{eq26}
%\small
{\bf{g}}_{optim}  = \mathop {\arg \min }\limits_{{\bf{g}}_i ,i \in \{ 1,2,...,F\} }dist({\bf{x}},{\bf{g}}_i )
\end{equation}

\par
\begin{figure}[htb!]
\begin{center}
  \scalebox{0.45}{\includegraphics[width=\textwidth]{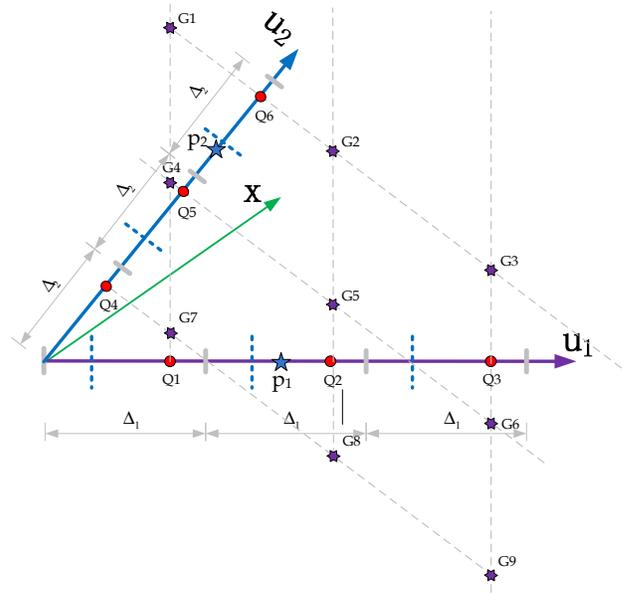}}
  \caption{Utilizing STDM-MW-Poptim to embed two message bits into one host vector $\bf{x}$, where $\bf{u}_1$ and $\bf{u}_2$ are the two projective vectors denoting the quantization directions. $Q_1$, $Q_2$ and $Q_3$ are the selected quantization points, corresponding to the search area $k=-1,0,1$ in \eqref{eq24}. These points form ${\bf{H}}_1$, the point-set of direction $\bf{u}_1$. $Q_4$, $Q_5$, $Q_6$ are the same ones.}\label{STDM_MW_OptimR}
\end{center}
\end{figure}

More specifically, Fig.\ref{STDM_MW_OptimR} gives an optimization example for STDM-MW-Poptim, which is the simple case of embedding two bits into one host vector. Three quantization points are selected in each directions, thus $3^2$ watermarked vectors ($G_1$,$G_2$,...,$G_9$) can be generated. The final optimized watermarked vector is $G_2$, the one that is closest to the host vector $\bf{x}$ among the nine candidate vectors.

\subsubsection{STDM-MW-Qoptim}
It has been observed that the goal of our optimization procedure is to find the optimal DM quantization point along each direction which makes the Euclidian distance between the optimized watermarked vector and the host vector is minimum. According to \eqref{eq27}, the Euclidian distance, $dis\_v$, can be expressed as follows,
\begin{equation}\label{eq47}
dis\_v =  \sqrt {{\bf{A}}^T {\bf{U}}_e {\bf{A}}}
\end{equation}
where ${\bf{A}} = ({\bf{Qp}}- {\bf{P}})$.
\par
Thus, the optimization procedure can be formulated as a constrained quadratic minimization problem that minimizes
\begin{equation}\label{eq28}
Y = {\bf{A}}^T {\bf{U}}_e {\bf{A}}
\end{equation}
subject to the constraint in the form of
\begin{equation}\label{eq29}
%\small
{\bf{A}} + {\bf{P}} \in {\bf{Qp\_S}}
\end{equation}
\par
To do the optimization, a part of elements in $\bf{Qp}$ are selected as the fixed elements, each of which is generated from quantizing the projection point using \eqref{eq3}, that is, the closest DM quantization point to the projection point. The other elements in $\bf{Qp}$ will be optimized to minimize $Y$ in \eqref{eq28}.
\par
Assuming the elements to be optimized in $\bf{Qp}$ are $Qp(o_1)$,$Qp(o_2)$,...,$Qp(o_t)$ and the elements to be fixed are $Qp(f_1)$,$Qp(f_2)$,...,$Qp(f_r)$, thus, in \eqref{eq28}, the corresponding elements to be optimized and fixed in $\bf{A}$, ${\bf{A}}={\bf{Qp}}- {\bf{P}}$, will be $A(o_1)$,$A(o_2)$,...,$A(o_t)$ and $A(f_1)$,$A(f_2)$,...,$A(f_r)$. By differentiating $Y$ with respect to each element to be optimized, and setting the derivatives to be zero, $t$ equations will be generated
\begin{equation}\label{eq30}
%\small
\frac{{\partial Y}}{{\partial A(o_i) }} = 0,\;i = 1,2,...,t
\end{equation}

\begin{equation}\label{eq31}
%\small
 \Rightarrow\;\sum\limits_{j = 1}^t {{\bf{U}}_{eo_io_j } A(o_j) }=\sum\limits_{k = 1}^r {{\bf{U}}_{eo_if_k } A(f_k)},\;i=1,2,...t
\end{equation}
\par
Solving \eqref{eq31}, $t$ optimized elements in $A$ are produced, consequently, the optimized elements in $\bf{Qp}$ can be generated, $Qp(o_i)=A(o_i)+P(o_i)$. Unfortunately, each $Qp(o_i)$ may not subject to the constraint \eqref{eq29}, in other words, $Qp(o_i)$ dose not belong to the set of quantization points of the $o_i$-th direction, set ${\bf{B}}_{o_i}$, $\{b|b=\beta_{o_i} \Delta_{o_i}  - d_{o_i}^{m_{0_i}},\;\beta_{o_i}  \in Z\}$. To satisfy this constraint, the final optimized $Qp(o_i)$ can be given by
\begin{equation}\label{eq32}
%\small
Qp(o_i)  = \mathop {\arg \min }\limits_{b_j ,b_j \in {\bf{B}}_{o_i} }dist(Qp(o_i),b_j)
\end{equation}
\par
Finally, vector $\bf{QoQp}$ is generated by assembling $Qp(o_i)$ and $Qp(o_f)$, and it can be used to substitute the vector $\bf{Qp}$ in \eqref{eq105}. The watermarked vector can be calculated by
\begin{equation}\label{eq33}
%\small
{\bf{g}}_i  = {\bf{x}} + {\bf{UU}}_{\bf{I}} ^{ - 1} ({\bf{QoQp}}_i  - {\bf{P}}),\;\;i=1,2,...,F
\end{equation}
where, assuming there are $n$ bits to be embedded in one host vector, in other words, there are $n$ given quantization directions for one host vector. Thus, there are $
F=\left( \begin{array}{l}
 r \\
 n \\
 \end{array} \right)
$ ways to choose $r$ elements from $\bf{Qp}$ (of length $n$) to play as the fixed elements. $F$ vectors $\bf{QoQp}$ are generated, and correspondingly, $F$ watermarked vectors $\bf{g}$ are produced.
\par
The final optimized watermarked vector ${\bf{g}}_{optim}$ is then given by judging which of these watermarked vectors produced in \eqref{eq33} has the minimum Euclidean distance to the host vector $\bf{x}$.
\begin{equation}\label{eq34}
%\small
{\bf{g}}_{optim}  = \mathop {\arg \min }\limits_{{\bf{g}}_i ,i \in \{ 1,2,...,F\} }dist({\bf{x}},{\bf{g}}_i )
\end{equation}

\par
\begin{figure}[htb!]
\begin{center}
  \scalebox{0.45}{\includegraphics[width=\textwidth]{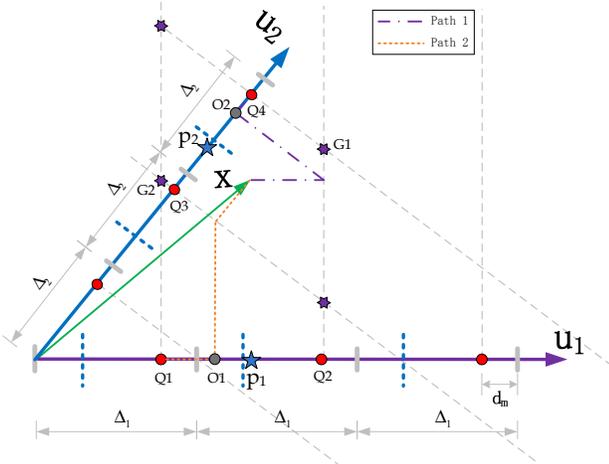}}
  \caption{Utilizing STDM-MW-Qoptim to embed two message bits into one host vector $\bf{x}$.}\label{STDM_MW_Optim}
\end{center}
\end{figure}

More specifically, Fig.\ref{STDM_MW_Optim} gives an optimization example for STDM-MW-Qoptim, which is the simple case of embedding two bits into one host vector. Thus, there are two elements in $\bf{Qp}$, $Qp(1)$ and $Qp(2)$, corresponding to the projection directions $\bf{u_1}$ and $\bf{u_2}$. If $Qp(1)$ is fixed, then $Qp(1)$ is equal to $Q_2$. Through \eqref{eq30}, actually Path 1 in Fig.\ref{STDM_MW_Optim}, the optimized point of $Qp(2)$ is $O_2$. Finally, according to \eqref{eq32}, $O_2$ is quantized to $Q_4$, and the corresponding watermarked vector is $G_1$. Correspondingly, if $Qp(2)$ is fixed, Path 2 is used to optimize $Qp(1)$, and $G_2$ is the corresponding watermarked vector. Comparing $G_1$ with $G_2$, the final optimal watermarked vector is $G_2$, the one that is closer to the host vector $\bf{x}$.

\section{Experimental Results and Analysis}

\begin{figure}[htb!]
  \centering
  \scalebox{0.35}{\includegraphics[width=\textwidth]{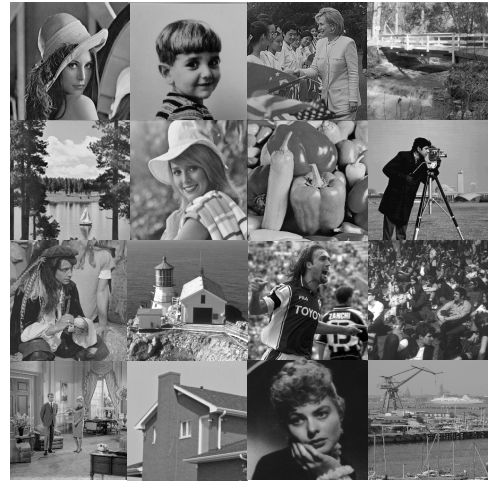}}
  \caption{Test images}\label{16image}
\end{figure}

To evaluate the performance of our proposed method, experiments are performed on standard images with size $256 \times 256$ as shown in Fig.\ref{16image}. And all the experiment data illustrated in the following section are the averaged ones.
\par

More specifically, for all the proposed algorithms, to be analyzed in the experiments, the $2^{nd}  - 8^{th}$ DCT coefficients, in zig-zag-scaned order, of each $8 \times 8$ block are used to form each host vector which is used to embed several message bits in it. The projective vectors and quantization steps are generated from the Gaussian distribution $\mathcal {N}(0,16)$ and $\mathcal {N}(f_g,4)$, respectively. $f_g$ is adjusted to ensure a given image fidelity.

\subsection{Experimental Test for the Efficiency of the Optimization Methods}
As mentioned above, to optimize the proposed multiple watermarking algorithm, two optimization methods, STDM-MW-Poptim and STDM-MW-Qoptim, are proposed to realize image fidelity improvement. To test their performance, 5 watermarks, with size $32\times32$, are embedded into the standard image. Meanwhile, the same quantization steps, dither signals and projective vectors are used for the two methods to compare their performance in PSNR\&CPU-time.
\begin{figure}[htb!]
\begin{center}
  \scalebox{0.5}{\includegraphics[width=\textwidth]{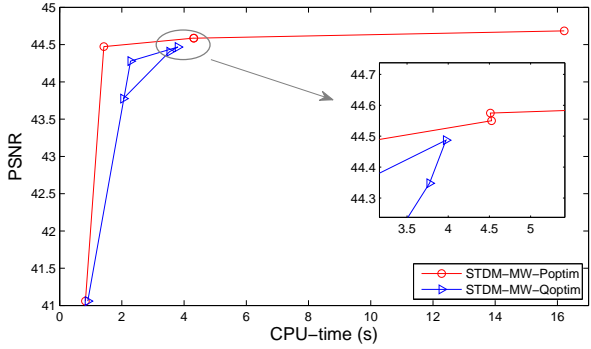}}
  \caption{PSNR Vs. CPU-time with different optimization parameters. The first point denotes embedding without optimization, the rest points are corresponding to the different search areas, $k=(0,1),(0,1,2),(-1,0,1),(-1,0,1,2)$ in \eqref{eq24}, for STDM-MW-Poptim and the fixed numbers, $r=4,3,2,1$ in \eqref{eq31}, for STDM-MW-Qoptim.}\label{psnr_cputime}
\end{center}
\end{figure}

As illustrated in Fig.\ref{psnr_cputime}, both of them have great performance in the improvement of the fidelity of the watermarked image. The image fidelity is promoted from 41dB to 44dB in PSNR, compared with the original embedding strategy.
\par
In STDM-MW-Poptim, along with the growth of the search area, it takes more time to realize the optimization, whereas, gives less contribution to the increase in PSNR. Taking a tradeoff between CPU-time and PSNR, $2^{nd}$ point, search area $k=0,1$, is the optimal one for five users in STDM-MW-Poptim.
\par
In STDM-MW-Qoptim, the CPU-time of $2^{nd}$\&$5^{th}$ point and $3^{rd}$\&$4^{th}$ point are almost the same. This is mainly due to the fact that the number of the watermarked vectors generated for one host vector, $F$ in \eqref{eq33}, are the same, because $
F=\left( \begin{array}{l}
 4 \\
 5 \\
 \end{array} \right)
 =\left( \begin{array}{l}
 1 \\
 5 \\
 \end{array} \right)
$ for $2^{nd}$\&$5^{th}$ point, and $
F=\left( \begin{array}{l}
 3 \\
 5 \\
 \end{array} \right)
 =\left( \begin{array}{l}
 2 \\
 5 \\
 \end{array} \right)
$ for $3^{rd}$\&$4^{th}$ point. Taking a tradeoff between CPU-time and PSNR, $4^{th}$ point, fix number $r=2$, is the optimal one for five users in STDM-MW-Qoptim.
\par
Comparing the two optimal points in the two methods, STDM-MW-Poptim has better performance due to less CPU-time and higher PSNR.
\par
Through experiments, for different numbers of users, it is found that $k=0,1$ and $r=floor(user\_number/2)$ are the appropriate optimization parameters for STDM-MW-Poptim and STDM-MW-Qoptim. And in what follows, these two parameters are used to implement the optimization.

\subsection{Experimental Test for the Proposed Methods in Robustness\&PSNR}
To test the impact of multiple watermarks embedding to the fidelity of the image, different numbers of watermarks are embedded into the image using the four proposed methods separately.
\par

\begin{figure}[htb!]
\begin{center}
  \scalebox{0.45}{\includegraphics[width=\textwidth]{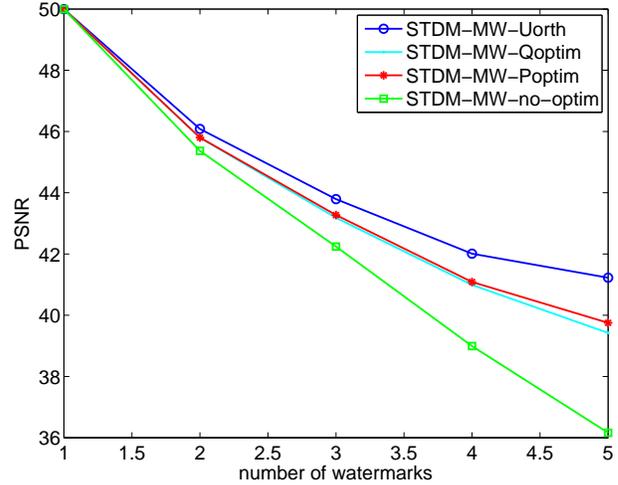}}
  \caption{PSNR Vs. Number of watermarks}\label{PSNR}
\end{center}
\end{figure}

As illustrated in Fig.\ref{PSNR}, along with the increase of the number of watermarks embedded, the quality of the images declines in vary degrees. STDM-MW-Uorth, which uses Gram-Schmidt orthogonalization to preprocess the projective vectors, has the superior image quality among these methods. This is mainly due to STDM-MW-Uorth is optimal in the case of orthogonal projective vectors. Unfortunately, in the general case that the projective vectors are not orthogonal, STDM-MW-no-optim, the quality of the watermarked image declines rapidly using the original embedding strategy without optimization. In contrast, if optimization is applied, e.g., STDM-MW-Poptim, the situation will be improved by a large scale, which is promoted by 1.03dB for 3 watermarks, 2.09dB for 4 watermarks, and 3.59dB for 5 watermarks.

\par
From another point of view, to evaluate the robustness of our proposed multiple watermarking methods, the test images are embedded into 3 watermarks, with size $32\times32$, under the uniform fidelity, a fixed PSNR of 42 dB. Meanwhile, four kinds of attacks, Gauss Noise, JPEG Compression, Salt\&Pepper Noise and Amplitude Scaling, are used to verify the performance of the schemes.

\par
\begin{figure*}[htb!]
\begin{center}
  \subfigure[] {\scalebox{0.46}{\includegraphics[width=\textwidth]{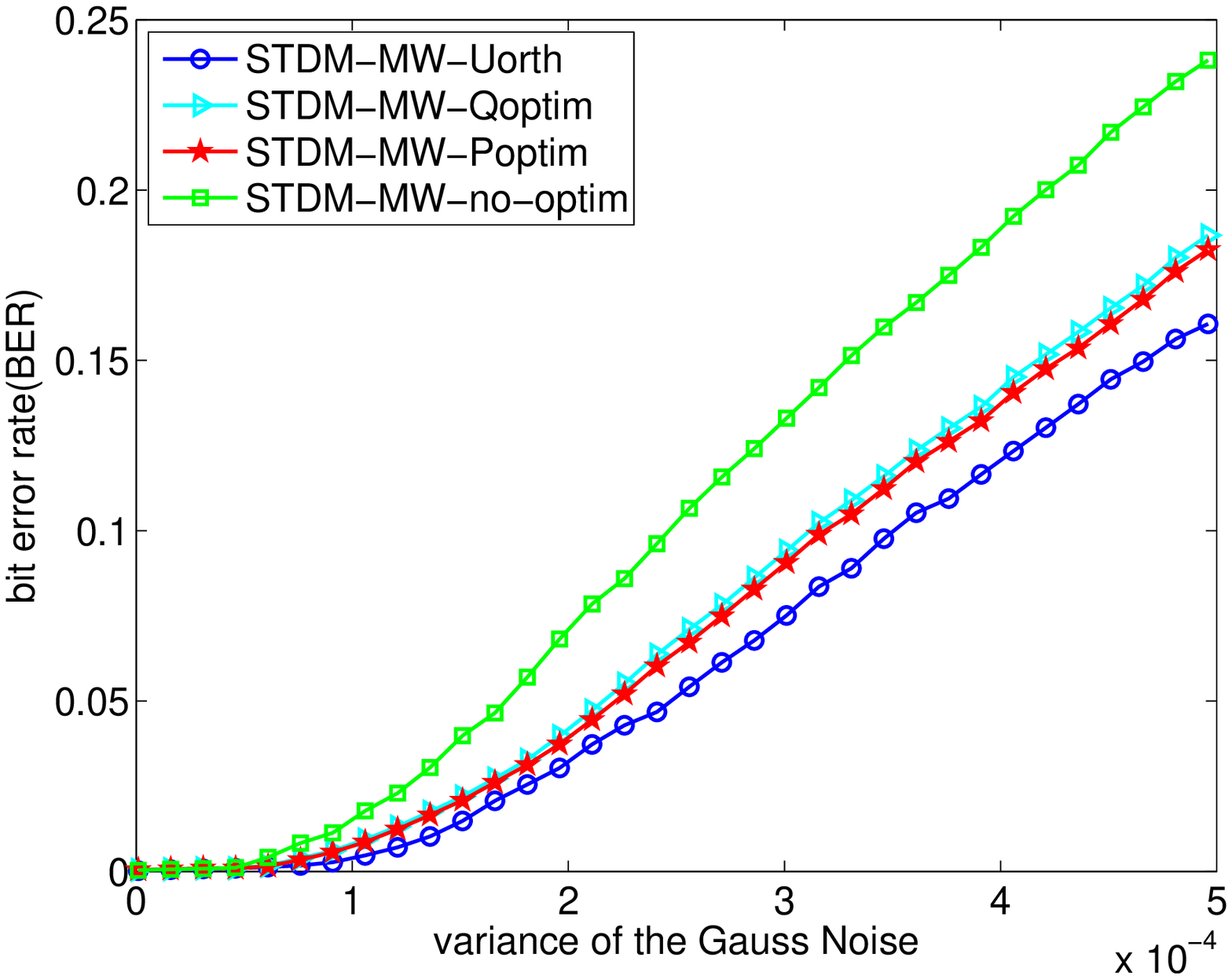}}}
  \subfigure[] {\scalebox{0.46}{\includegraphics[width=\textwidth]{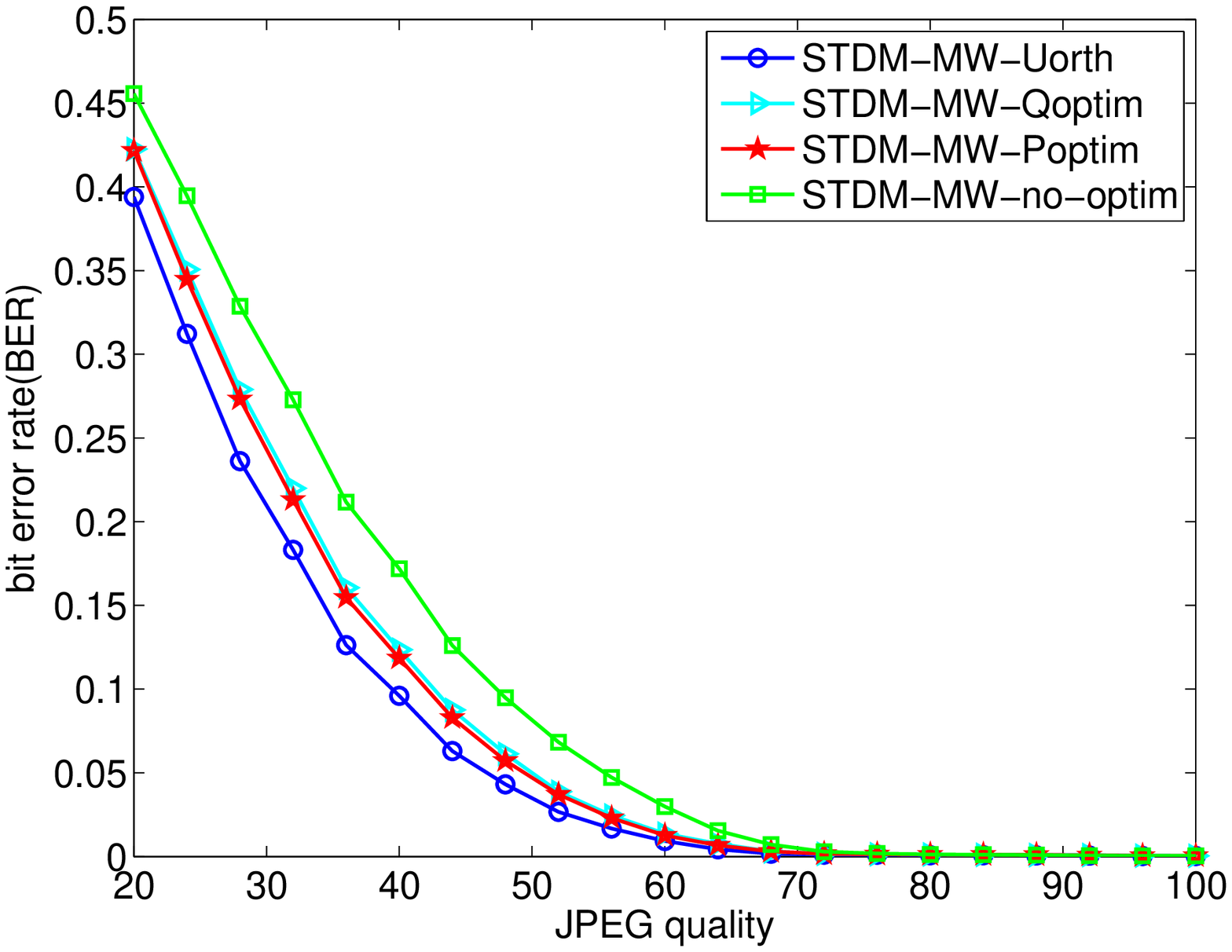}}}
  \subfigure[] {\scalebox{0.46}{\includegraphics[width=\textwidth]{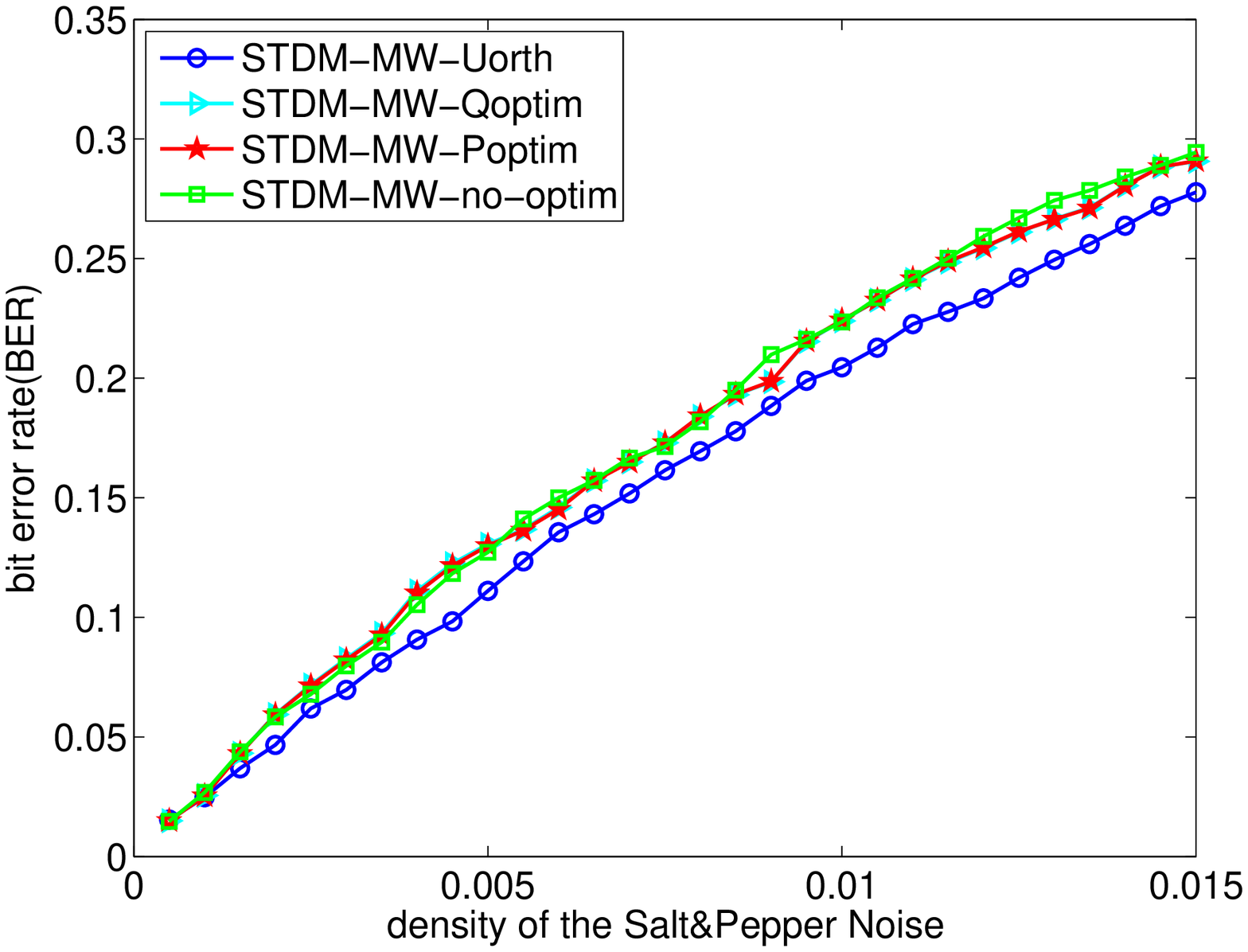}}}
  \subfigure[] {\scalebox{0.46}{\includegraphics[width=\textwidth]{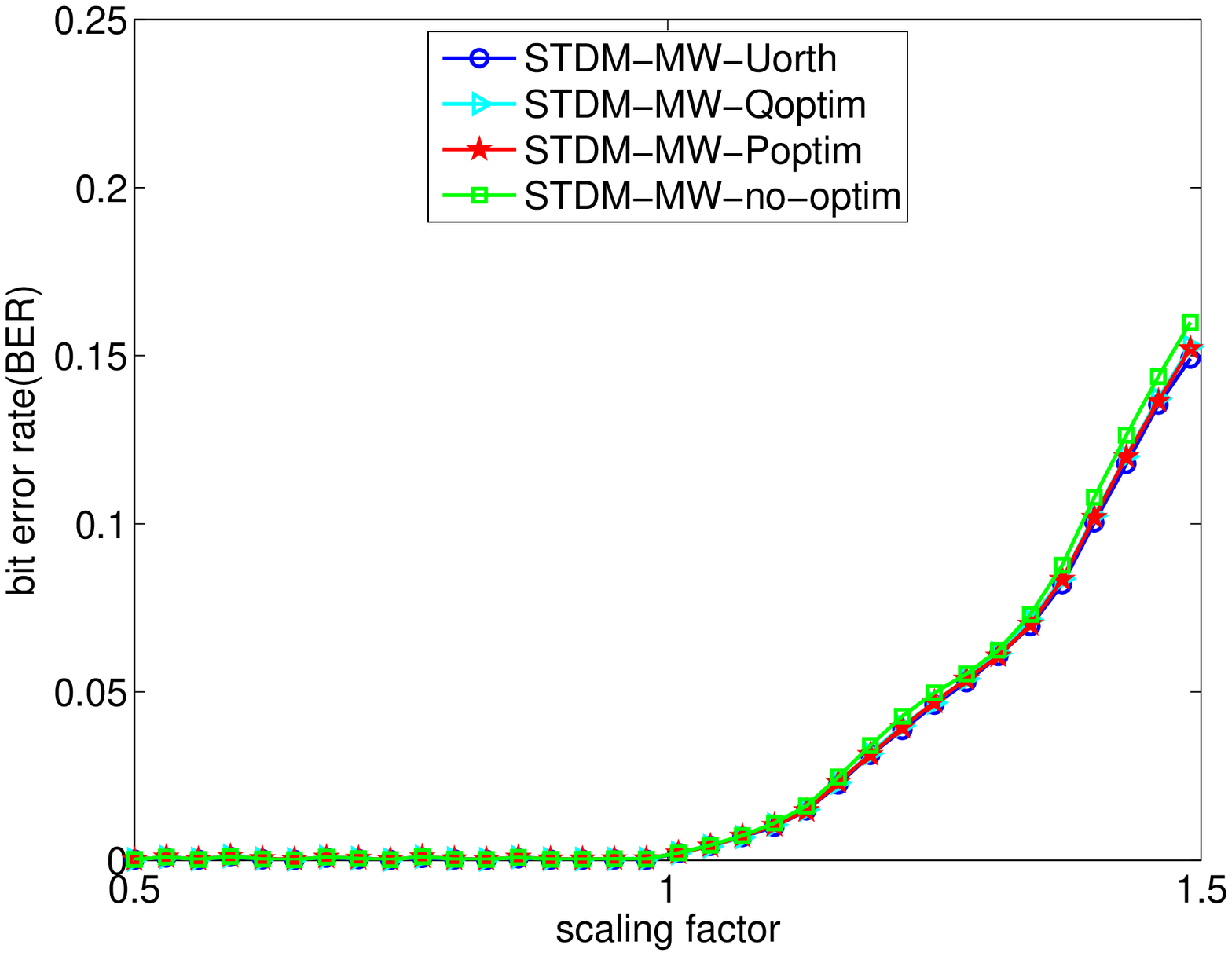}}}
%%  \scalebox{0.8}{\includegraphics[width=\textwidth]{inter4totle.eps}}
  \caption{BER vs. (a) Gaussian Noise, (b) JPEG, (c) Salt\&Pepper Noise and (d) Amplitude Scaling}\label{inner_compare}
\end{center}
\end{figure*}

As illustrated in Fig.\ref{inner_compare}, we test four versions, STDM-MW-no-optim, STDM-MW-Poptim, STDM-MW-Qoptim and STDM-MW-Uorth. And we use the average detection score, measured in bit error rate (BER), to analyze the performance, and each curve is the average BER of the three detected watermarks.

\par
As we expected, according to Fig.\ref{inner_compare}.(d), all the proposed schemes do have good performance in amplitude scaling. The rise of BER in scale $\beta  \ge 1.2$ is mainly due to the ``cutoff distortion", that is, some pixels of the image are already quite huge and will be cut off to the maximum allowed value when there is an scaling. In this case, the pixels will not scale linearly with the scaling factor while the quantization step-sizes still scale linearly as usual. Thus, experimental performance on bright images will have a worse robustness in this scale.
\par
With regard to other attacks, both STDM-MW-Poptim and STDM-MW-Qoptim have better robustness against Gauss noise (Fig.\ref{inner_compare}.(a)) and JPEG compression (Fig.\ref{inner_compare}.(b)) compared with STDM-MW-no-optim. This mainly due to the fact that the optimization procedures can improve the fidelity of the watermarked image, as shown in Fig.\ref{PSNR}, in other words, the embedding strength used in them could be relatively increased while ensuring the given fidelity.
\par
Although the STDM-MW-Uorth is the best performed one, it is not suitable for the applications where independent detection is required, because all the projective vectors of each users must be gained in the detector to perform Gram-Schmidt orthogonalization before the detecting procedure. Thus, referencing to section VI-A, STDM-MW-Poptim is the optimal one to play as the multiple watermarks embedding strategy in the sense of higher robustness, less CPU-time and for general applications.

\begin{figure*}[htb!]
\begin{center}
  \subfigure[] {\scalebox{0.46}{\includegraphics[width=\textwidth]{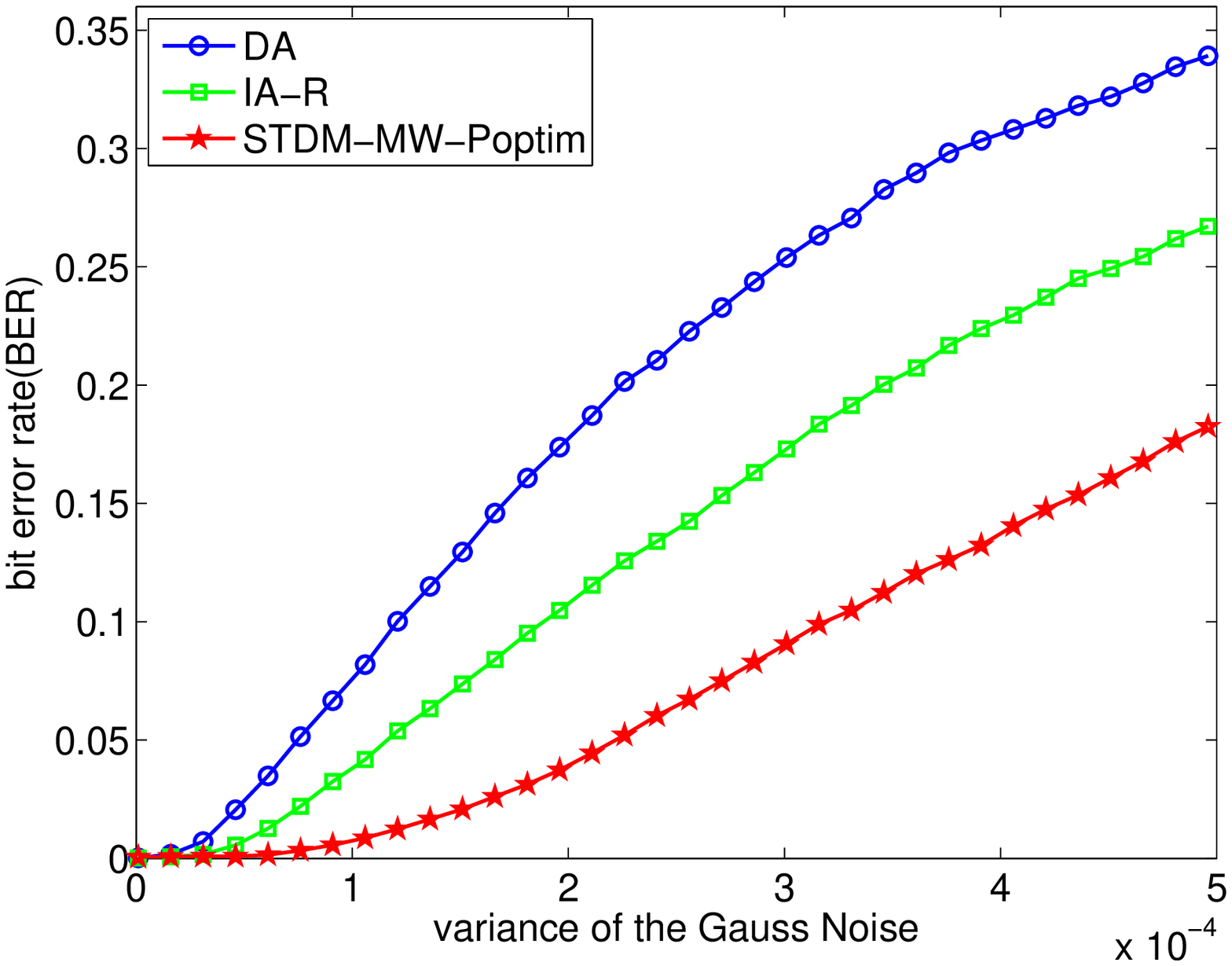}}}
  \subfigure[] {\scalebox{0.46}{\includegraphics[width=\textwidth]{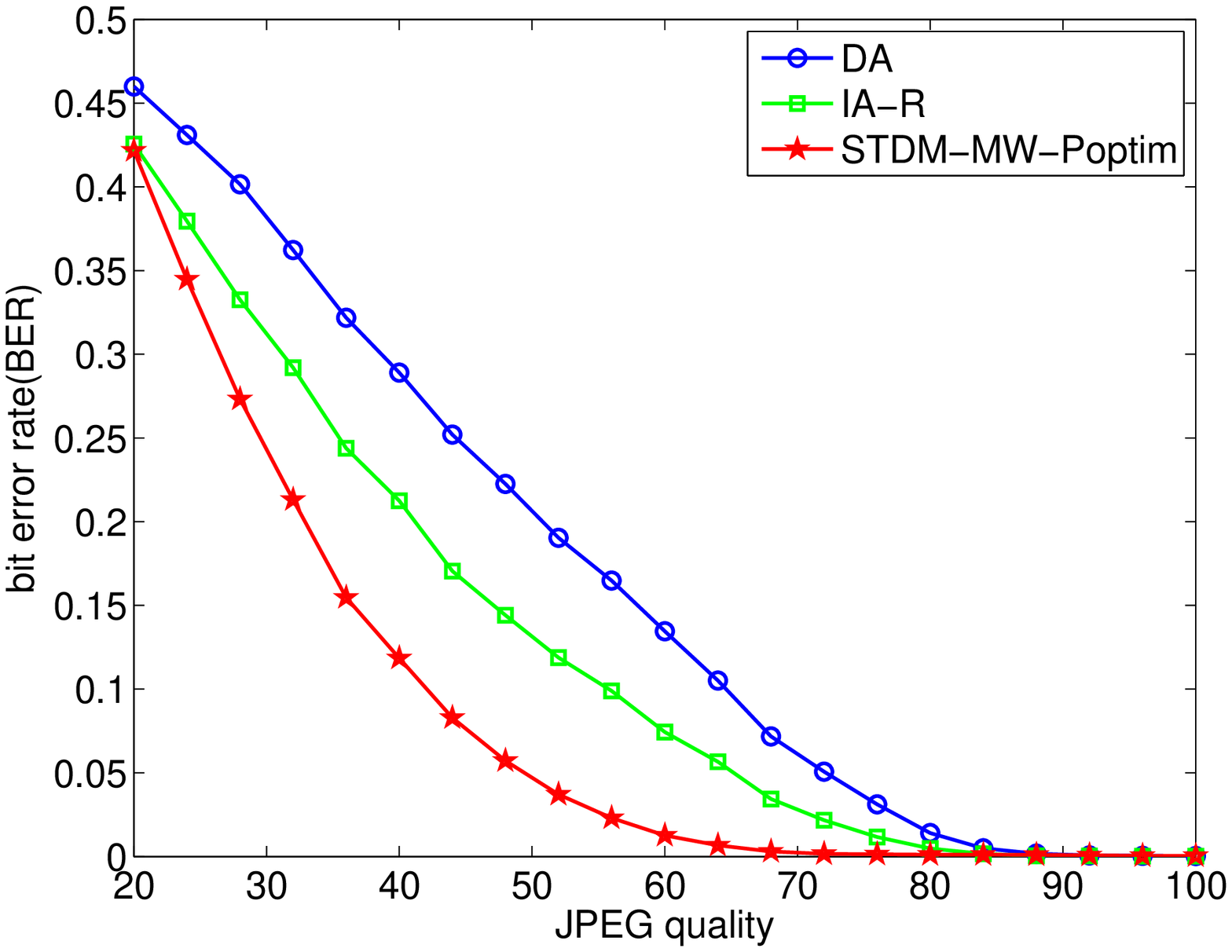}}}
  \subfigure[] {\scalebox{0.46}{\includegraphics[width=\textwidth]{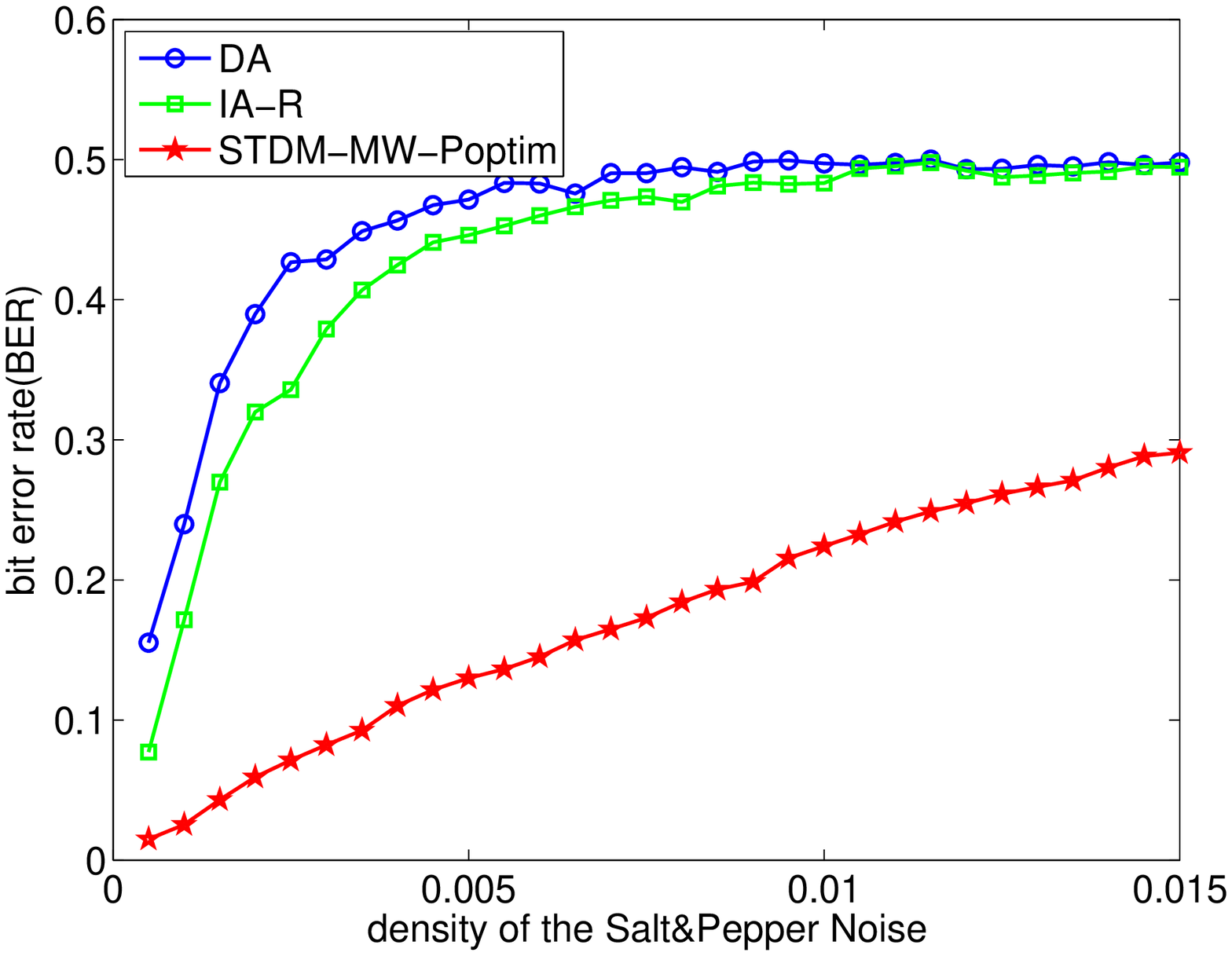}}}
  \subfigure[] {\scalebox{0.46}{\includegraphics[width=\textwidth]{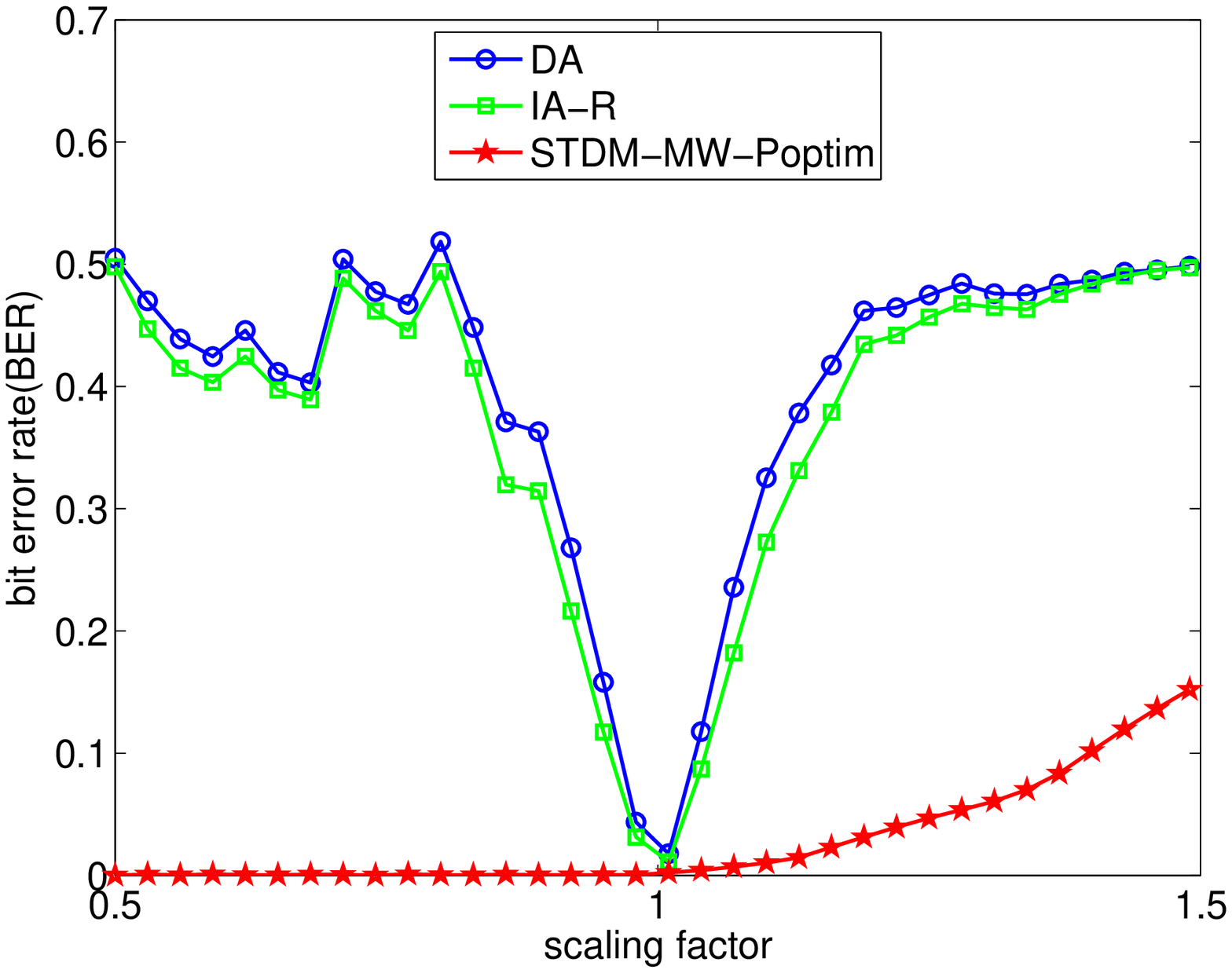}}}
%%  \scalebox{0.8}{\includegraphics[width=\textwidth]{inter4totle.eps}}
  \caption{BER vs. (a) Gaussian Noise, (b) JPEG, (c) Salt\&Pepper Noise and (d) Amplitude Scaling}\label{inter_compare}
\end{center}
\end{figure*}

\begin{figure*}[htb!]
\begin{center}
  \scalebox{0.95}{\includegraphics[width=\textwidth]{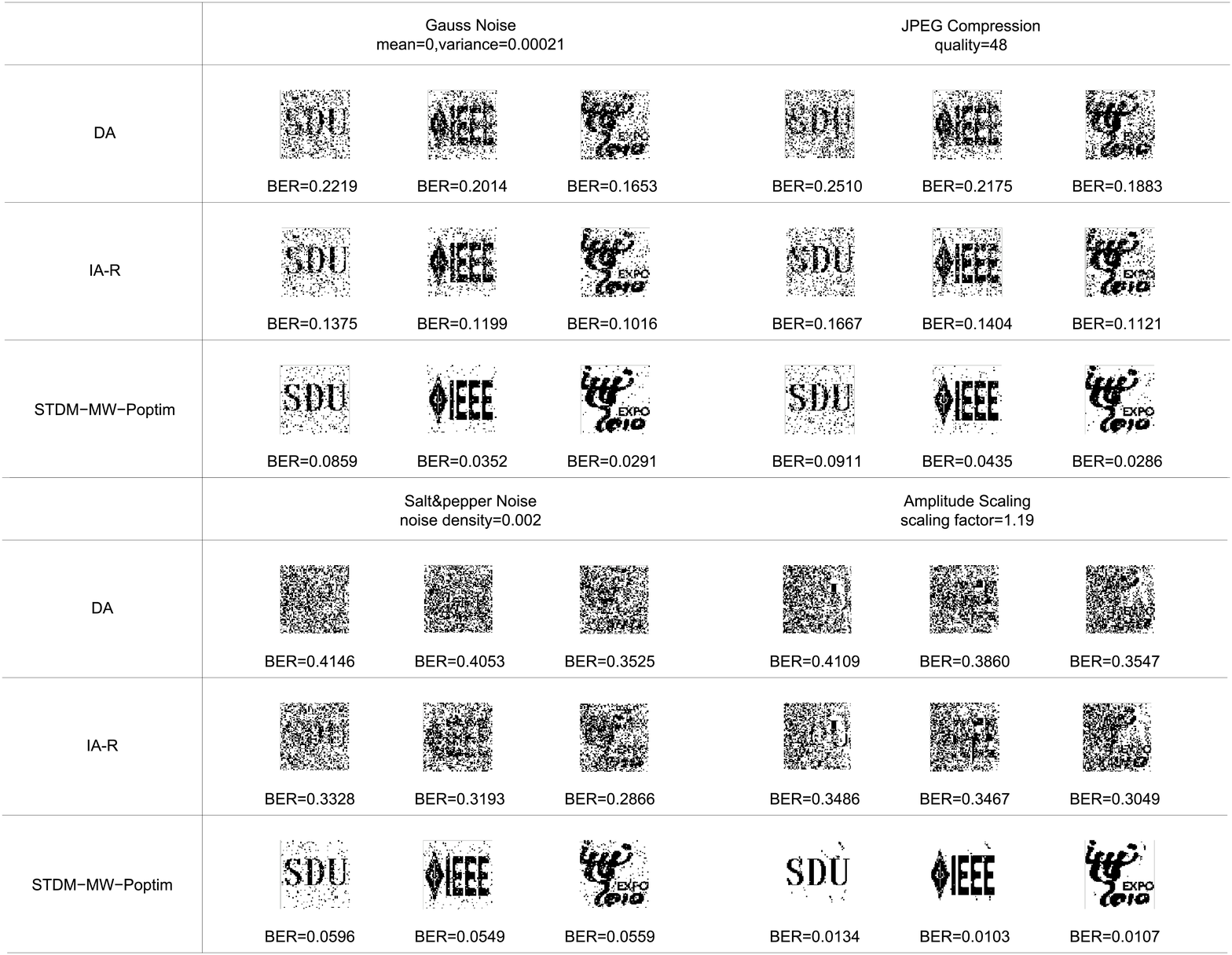}}
  \caption{Watermark show}\label{watermark_show}
\end{center}
\end{figure*}

\subsection{Comparison with the Pioneering Multiple Watermarking Algorithms}
To give an objective analysis of the performance of the proposed method, the optimal one of our proposed schemes, STDM-MW-Poptim, is picked up to be compared with the pioneering multiple watermarking algorithms, DA and IA-R in \cite{MW_wang2003}. Both of them can embed multiple watermarks into the same image area, and each watermark can be detected independently, like ours. To correspond with the original paper, the parameters used for them are  identical, the keys K is generated from Gaussian distribution $\mathcal {N}(0,16)$ and the first 10\% of the DCT AC coefficients are used to form the host vector, meanwhile, 3 watermarks are embedded into the standard images, the same as ours. Note that, the mean of the keys D is modified to meet the uniform image fidelity, 42dB in PSNR. The BER curves are illustrated in Fig.\ref{inter_compare}, meanwhile, to show the subjective visual effect, the detected watermarks corresponding to different conditions are given in Fig.\ref{watermark_show}.
\par
As illustrated in Fig.\ref{inter_compare}.(d), because DA and IA-R do not take the amplitude scaling attack into account, they cannot resist the image process which scales the amplitude of the pixels. In contrast, STDM-MW-Poptim has great advantage in this field,
\par
In robustness to random noise and JPEG Compression, Fig.\ref{inter_compare}.(a)(b)(c), our proposed scheme outperforms others significantly, especially in Salt\&Pepper Noise attack, the performance is almost improved by 70\%. Such superior performance is attributed to the exploitation of the great robustness of the original STDM in single watermarking. In addition, the optimization strategy can provide a significant improvement in image fidelity, in other words, the embedding strength used in our scheme could be relatively increased while ensuring the given fidelity.

\section{Application Discussion and Extension}
As mentioned above, the proposed multiple watermarking algorithm has the feature that it can embed multiple watermarks into the same area and the same transform domain of one image, meanwhile, the embedded watermarks can be extracted independently and blindly in the detector without any interference. To this end, it may own some potential interesting applications.

\subsection{Coauthor Copyright Certification}
In the field of copyright management, one common scenario is that a number of authors who have co-designed an image need separate certification for each of them. This can be fulfilled by the proposed algorithm, STDM-MW-Poptim, in which the embedded watermarks (certifications for each author) can be extracted independently and blindly in the detector. Every author can use his/her own key set, $STEP\_KEY$, $U\_KEY$ and $Dither\_KEY$, to extract his/her own watermark, by which the copyright of each author can be certificated independently.

\subsection{Secret Related Area}
A more interesting feature of STDM-MW-Poptim is that the detecting procedure of each watermark is independent with each other. More importantly, the receiver even dose not know how many watermarks are exactly embedded, i.e., one receiver cannot perceive the exist of other hidden information without the notification from the embedder. This is due to the fact that in terms of each receiver, the detecting procedure is exactly the same as STDM, which is deemed as a single watermarking algorithm. This interesting feature would cause the gloss to the receiver that the watermark he/she has extracted is the only information hidden in the image, and this gloss may provide a key cover for the protection of the true secret information.

\subsection{Image History Management}
In some applications such as medical image management, it is desirable to acquire the history of a medical image from the patient through the various laboratories and physicians, e.g., directly detecting from the image who is the creator, who has access to the data after its creation. This can be realized by sequentially embedding each user's digital signature into the image during each stage of its circulation.
\par
Inspired by \cite{MW_wang_sequential}, we can utilize the special case of our proposed algorithm, multiple watermarking using orthogonal projective vectors, STDM-MW-Uorth, combined with STDM-MW-Poptim to fulfill this application.
\par
\begin{figure*}[htb!]
\begin{center}
  \scalebox{0.94}{\includegraphics[width=\textwidth]{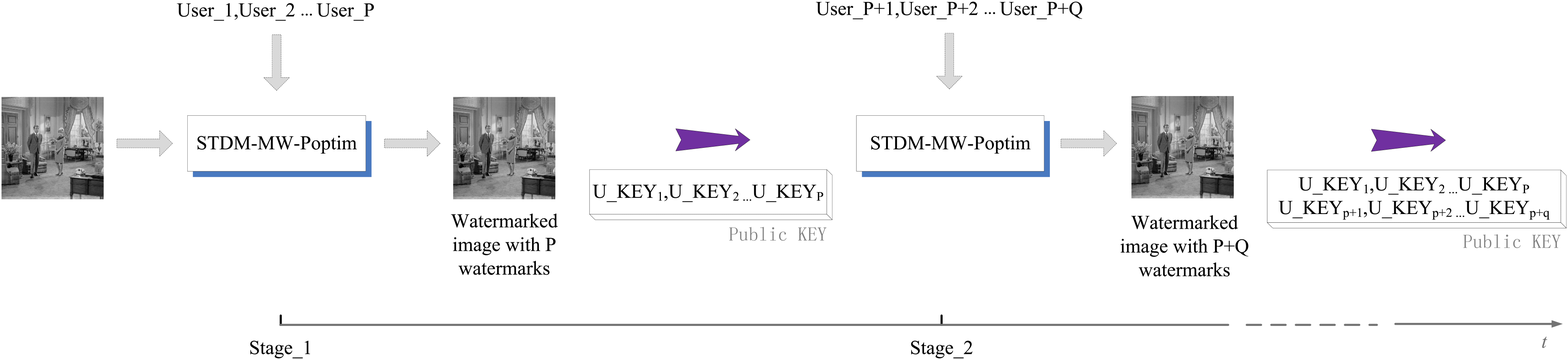}}
  \caption{Sequential Multiple Watermarks Embedding}\label{MW_Sequential}
\end{center}
\end{figure*}

As illustrated in Fig.\ref{MW_Sequential}, if Q additional watermarks are desired to be embedded into the watermarked image with P watermarks embedded, we must guarantee that these additional watermarks must not interfere with the former embedded watermarks. To realize this, we apply the idea of STDM-MW-Uorth, using projective vectors that are orthogonal to the ones of the former embedded watermarks.

\par
To embed Q additional watermarks simultaneously for the coming Q users, the watermarked image with P watermarks embedded as well as a public key set (the former users' $U\_KEY$) are needed. Then, the projective vector ${\bf{u}}_i$ produced by each new user will be preprocessed by Gram-Schmidt orthogonalization.
\begin{equation}\label{eq37}
%\small
{\bf{u}}^{orth} _i  = {\bf{u}}_i  - \sum\limits_{j = 1}^P {proj({\bf{u}}_i ,{\bf{k}}_j ) \cdot \frac{{{\bf{k}}_j }}{{\left\| {{\bf{k}}_j } \right\|_2 }}\;\;,i = 1,2,...,Q}
\end{equation}

Finally, based on these preprocessed projective vectors, ${\bf{u}}^{orth} _1$,${\bf{u}}^{orth} _2$,...,${\bf{u}}^{orth} _Q$,  Q additional watermarks can be simultaneously embedded into the watermarked image using STDM-MW-Poptim without any interference.
\par
One step further, if all the watermarks are desired to be embedded into the image one by one, this case is equivalent to STDM-MW-Uorth.
\par
In this way, we can embed multiple watermarks into the image sequentially to realize image history management. Compared with \cite{MW_Tracing}, which is based on \cite{SW_07MW_Tracing_base1,SW_07MW_Tracing_base2}, an additional management for the public key set is needed in our scheme. Nevertheless, to detect the watermark, \cite{MW_Tracing} must acquire the knowledge of the content of the original embedded watermark to implement correlation detection and can only judge whether there exists the given watermark. This feature may somehow constrain its application area.

\section{Conclusions}
In this paper, a novel multiple watermarking algorithm is presented which initially extend the STDM, a single watermarking algorithm, to the field of multiple watermarking application. It can embed multiple watermarks into the same area and the same transform domain of one image; meanwhile, the embedded watermarks can be extracted independently and blindly in the detector without any interference. Moreover, through investigating the properties of the DM quantizer and the proposed multiple watermarks embedding strategy, two optimization methods are presented to improve the fidelity of the watermarked image. Experimental results indicate that the optimization procedure can significantly improve the quality of the watermarked image, meanwhile, the more watermarks embedded the more quality improvements can be gained. Finally, to enhance the application flexibility, an application extension of our algorithm is proposed, which can sequentially embed multiple watermarks into the image during each stage of its circulation, thereby realizing image history management. In general, compared with the pioneering multiple watermarking algorithms, the proposed scheme owns more flexibility in practical application and is more robust against distortion due to basic operations such as random noise, JPEG compression and valumetric scaling.

\appendices
\section{}
Referencing \eqref{eq17}, to make it tenable, the matrix ${\bf{U}}_{_{\bf{I}} }$ must be reversible. As ${\bf{U}}_{_{\bf{I}} }$ is an n-by-n matrix, thus
\[
rank({\bf{U}}_{_{\bf{I}} })=n
\]
Referencing \eqref{eq16}, ${\bf{U}}_{\bf{I}}  = \Lambda _U {\bf{U}^T}{\bf{U}}$, thus
\[
rank({\bf{U}}_{\bf{I}} ) \le \min \{ rank({\bf{U}}),rank\{ {\bf{U'}}\} \}
\]
Consequently, we have $rank({\bf{U}}) \ge n$, and reference \eqref{eq14}, ${\bf{U}}$ is an L-by-n matrix, thus
\[
\left\{ \begin{array}{l}
 rank({\bf{U}}) = n \\
 L \ge n \\
 \end{array} \right.
\]
where L denotes the length of the host vector ${\bf{x}}$.

\section{}
Consider $X$ and $X'$ represent the original image and the watermarked one in the space domain. And referencing the DCT transformation, we have
\[
%\small
Y = AXA^{\rm T},\;\;Y' = AX'A^{\rm T}
\]
where $Y$ and $Y'$ are the coefficients in DCT domain.
\par
Then, the MSE between the original image and the watermarked image can be written by
\[
%\small
\begin{array}{l}
 MSE = \frac{1}{{mn}}\sum\limits_{i = 1}^{m } {\sum\limits_{j = 1}^{n } {[X(i,j) - X'(i,j)]^2 } }  \\
  \;\;\;\;\;\;\;\;\;\;=\frac{1}{{mn}}\left\| {X - X'} \right\|_F^2  \\
  \;\;\;\;\;\;\;\;\;\;=\frac{1}{{mn}}\left\| {A^{\rm T} YA - A^{\rm T} Y'A} \right\|_F^2  \\
  \;\;\;\;\;\;\;\;\;\;=\frac{1}{{mn}}\left\| {A^{\rm T} (Y-Y')A} \right\|_F^2  \\
 \end{array}
\]
where $\left\| Q \right\|_F$ denotes the Frobenius norm of the matrix $Q$, in view of $\left\| Q \right\|_F  \buildrel \Delta \over = (\sum\limits_{i = 1}^m {\sum\limits_{j = 1}^n {(Q(i,j))^2 )^{1/2} } }  = (tr(Q^T Q))^{1/2}$,
\[
%\small
 MSE =\frac{1}{{mn}}\times tr((A^{\rm T} (Y-Y')A)^T(A^{\rm T} (Y-Y')A)) \\
\]
Considering $A^T=A^{-1}$ in the DCT transformation, thus
\[
%\small
\begin{array}{l}
 MSE =\frac{1}{{mn}}\times  tr(A^{\rm T} (Y-Y')^T(Y-Y')A) \\
  \;\;\;\;\;\;\;\;\;\;=\frac{1}{{mn}}\times  tr((Y-Y')^T(Y-Y')) \\
  \;\;\;\;\;\;\;\;\;\;=\frac{1}{{mn}}\left\| {Y - Y'} \right\|_F^2
 \end{array}
\]
When $Y$ and $Y'$ are grouped into 1-dimension vectors, we have
\[
%\small
 MSE =\frac{1}{{N}}\left\| {C - C'} \right\|_2^2
\]
where $N=mn$, denotes the total number of the elements in the vector.

%% trigger a \newpage just before the given reference
%% number - used to balance the columns on the last page
%% adjust value as needed - may need to be readjusted if
%% the document is modified later
%%\IEEEtriggeratref{8}
%% The "triggered" command can be changed if desired:
%%\IEEEtriggercmd{\enlargethispage{-5in}}
%
%% references section
%
%% can use a bibliography generated by BibTeX as a .bbl file
%% BibTeX documentation can be easily obtained at:
%% http://www.ctan.org/tex-archive/biblio/bibtex/contrib/doc/
%% The IEEEtran BibTeX style support page is at:
%% http://www.michaelshell.org/tex/ieeetran/bibtex/
\bibliographystyle{IEEEtran}
\bibliography{xinba}
%%
%% <OR> manually copy in the resultant .bbl file
%% set second argument of \begin to the number of references
%% (used to reserve space for the reference number labels box)
%\vfill
%

\end{document}